\documentclass[aps,prx,notitlepage,superscriptaddress,longbibliography,twocolumn,nofootinbib,floatfix,10pt]{revtex4-2} 
\usepackage[inkscapelatex=false]{svg}
\usepackage[utf8]{inputenc}
\usepackage{graphicx}
\usepackage{hyperref}
\usepackage{dsfont}
\usepackage{tikz,tikz-3dplot}
\usetikzlibrary{decorations.pathreplacing}

\usepackage{lipsum}
\usepackage{leftindex}
\usepackage{amsmath,amsthm,amssymb,mathtools}
\usepackage{multirow}
\usepackage{braket}
\usepackage{bbm,bm}
\usepackage[bb=boondox,scr=boondox]{mathalfa}
\usepackage{enumitem}

\DeclareFontFamily{U}{matha}{\hyphenchar\font45}
\DeclareFontShape{U}{matha}{m}{n}{
      <5> <6> <7> <8> <9> <10> gen * matha
      <10.95> matha10 <12> <14.4> <17.28> <20.74> <24.88> matha12
      }{}
\DeclareSymbolFont{matha}{U}{matha}{m}{n}
\DeclareMathSymbol{\muparrow}{3}{matha}{"D2}
\DeclareMathSymbol{\mdownarrow}{3}{matha}{"D3}
\DeclareMathSymbol{\mupdownarrow}{3}{matha}{"D9}

\DeclareFontFamily{U}{mathb}{\hyphenchar\font45}
\DeclareFontShape{U}{mathb}{m}{n}{
      <5> <6> <7> <8> <9> <10> gen * mathb
      <10.95> mathb10 <12> <14.4> <17.28> <20.74> <24.88> mathb12
      }{}
\DeclareSymbolFont{mathb}{U}{mathb}{m}{n}
\DeclareMathSymbol{\mdownuparrows}{3}{mathb}{"D7}

\usepackage{makecell}

\usepackage{float}

\definecolor{dgreen}{rgb}{0,0.666,0}
\definecolor{dorange}{rgb}{0.666,.333,0}

\newcommand{\ie}{{\it i.e.},\ }

\newcommand{\id}{\mathbb{1}}
\newcommand{\Tr}{\operatorname{Tr}}
\newcommand{\ii}{\operatorname{i}}
\newcommand{\ee}{\operatorname{e}}

\newcommand{\Ai}{\operatorname{Ai}}
\newcommand{\Bi}{\operatorname{Bi}}

\makeatletter
\pgfmathdeclarefunction{erf}{1}{%
  \begingroup
    \pgfmathparse{#1 > 0 ? 1 : -1}%
    \edef\sign{\pgfmathresult}%
    \pgfmathparse{abs(#1)}%
    \edef\x{\pgfmathresult}%
    \pgfmathparse{1/(1+0.3275911*\x)}%
    \edef\t{\pgfmathresult}%
    \pgfmathparse{%
      1 - (((((1.061405429*\t -1.453152027)*\t) + 1.421413741)*\t 
      -0.284496736)*\t + 0.254829592)*\t*exp(-(\x*\x))}%
    \edef\y{\pgfmathresult}%
    \pgfmathparse{(\sign)*\y}%
    \pgfmath@smuggleone\pgfmathresult%
  \endgroup
}
\makeatother

\newtheorem{definition}{Definition}
\newtheorem{proposition}{Proposition}

\newtheorem{example}{Example}

\begin{document}

\title{Adiabatic vacua from linear complex structures}

\author{Eugenio Bianchi}
\email{ebianchi@psu.edu}
\affiliation{Department of Physics, The Pennsylvania State University, University Park, PA 16802, USA}
\affiliation{Institute for Gravitation and the Cosmos, The Pennsylvania State University, University Park, PA 16802, USA}

\author{Yusuf Ghelem}
\email{yaghelem@student.unimelb.edu.au}
\affiliation{School of Mathematics and Statistics, The University of Melbourne, Parkville, VIC 3010, Australia}

\author{Lucas Hackl}
\email{lucas.hackl@unimelb.edu.au}
\affiliation{School of Mathematics and Statistics, The University of Melbourne, Parkville, VIC 3010, Australia}
\affiliation{School of Physics, The University of Melbourne, Parkville, VIC 3010, Australia}

\begin{abstract}
Adiabatic vacua play a central role in quantum fields in cosmological spacetimes, where they serve as distinguished initial conditions and as reference states for the renormalization of observables. In this paper we introduce new methods based on linear complex structures which provide a powerful tool for determining adiabatic vacua. The new methods generalize both the standard WKB appoach and the Lewis-Riesenfeld invariants, and allow us to study the problem of many coupled bosonic degrees of freedom with general quadratic time-dependent Hamiltonian. We show that the adiabatic number operator and the adiabatic vacuum of finite order can be expressed in terms of the adiabatic complex structure of the same order. We compare our results to standard techniques which apply only to a single degree of freedom, and comment on its applicability to problems in quantum fields in cosmological spacetimes, many-body systems and quantum thermodynamics, where the Hamiltonian is time dependent with slowly-changing parameters.

\end{abstract}

\maketitle

\section{Introduction}
Adiabatic vacua play a central role in the theory of quantum fields in cosmological spacetimes \cite{birrell1984quantum,fulling1989aspects,Parker:2009uva}, where they serve as initial conditions for quantum  perturbations \cite{bunch1978quantum,mukhanov1992theory,Bianchi:2024qyp} and as reference states for the renormalization of the energy-momentum tensor via adiabatic subtraction \cite{parker1974adiabatic,birrell1978application,agullo2015preferred}. The standard construction introduced by Parker in \cite{Parker:1968mv,Parker:1969au,Parker:2012at} uses the Wentzel-Kramer-Brillouin (WKB) method \cite{messiah2014quantum,bender2013advanced,white2010asymptotic,Winitzki_2005} applied individually to each decoupled Fourier mode of the quantum field. In this paper, we introduce a new construction of adiabatic vacua based on the notion of adiabatic initial conditions for the linear complex structure and its associated adiabatic number operator \cite{GaussianStatesFromKaehler}. This formulation generalizes both the WKB method and the notion of the Lewis-Riesenfeld adiabatic invariants~\cite{lewis1969exact}, allowing us to go beyond the single decoupled oscillator: We derive a general formula for adiabatic vacua of quantum systems described by a quadratic time-dependent Hamiltonian which couples $d$ bosonic degrees of freedom.


\medskip

The simplest example where adiabatic vacua arise is the case of an oscillator with time-dependent frequency $\omega(t)$. The Hamiltonian of the system is $\hat{H}(t)=\frac{1}{2}(\hat{p}^2+\omega(t)^2\,\hat{q}^2)$. At a given reference time $t_0$, we can consider the instantaneous vacuum $|0,t_0\rangle$, defined as the lowest-energy eigenstate of $\hat{H}(t_0)$. As the Hamiltonian at two different times does not commute, $[\hat{H}(t_0),\hat{H}(t_1)]\neq 0$, in general, under unitary time evolution the instantaneous vacuum is mapped to an excited state: The vacuum evolves into a superposition of instantaneous energy eigenstates at a later time, providing the mechanism  for cosmological particle production \cite{Ford:2021syk}. Remarkably, under the assumption that the frequency $\omega(t)$ changes slowly, the adiabatic theorem guaranties that a system prepared in the instantaneous vacuum remains approximately in its instantaneous vacuum at subsequent times \cite{messiah2014quantum}. This adiabatic-following phenomenon has important applications in areas ranging from quantum thermodynamics~\cite{gemmer2009quantum} to quantum computing~\cite{albash2018adiabatic}. The adiabatic vacuum $|0_{\bar{n}},t_0\rangle$ of order $\bar{n}$ improves the adiabatic-following condition by including higher order corrections in terms of the slowness parameters of the Hamiltonian. More explicitly, higher order corrections are obtained by introducing a dependence on the first $\bar{n}$ time derivatives of the frequency $\omega(t)$, instead of its instantaneous value alone. In this way, the construction of the adiabatic vacuum is teleological: It knows the protocol for the change of the frequency in advance and selects a state that foresees the change in the Hamiltonian to minimize its future non-adiabaticity, at least in a small neighborhood of the initial time $t_0$.

\medskip

The standard construction of adiabatic vacua is based on the WKB or phase integral method \cite{messiah2014quantum,bender2013advanced,white2010asymptotic,Winitzki_2005} which allows one to obtain approximate solutions of the time-dependent equations of motion of the position of the oscillator. The WKB frequency $W(t)$ provides an approximate solution of a non-linear differential equation, obtained as an expansion around the instantaneous frequency, $W(t)\approx \omega(t)$, with corrections that depend on the first $\bar{n}$ time derivatives of the frequency, $\dot{\omega}(t), \ddot{\omega}(t), \ldots$, that defines the Hamiltonian $\hat{H}(t)$. While the solution $W(t)$ is only approximate, the values of $W(t_0)$ and $\dot{W}(t_0)$ can be taken as an exact definition of the initial conditions for an adiabatic solution of order $\bar{n}$ \cite{agullo2015preferred}. This choice selects a specific notion of positive frequency solution \cite{Ashtekar:1975zn,Ashtekar:1980yw,Wald:1995yp,Derezinski:2013dra},
\begin{equation}
v(t)= \tfrac{1}{\sqrt{2W(t)}}\ee^{-\ii \int W(t) dt}\;,
\end{equation}
and therefore of decomposition of the Heisenberg position operator into creation and annihilation operators, $\hat{q}_H(t)=v(t)\,\hat{a}+v^*(t)\,\hat{a}^\dagger$. The adiabatic vacuum of order $\bar{n}$ is then defined as the state satisfying the annihilation condition $\hat{a}|0_{\bar{n}},t_0\rangle=0$ at the reference time $t_0$. The WKB method is most often used in the study of quantum fields in cosmological spatimes \cite{birrell1984quantum,fulling1989aspects,Parker:2009uva}.

There is also an alternative construction of adiabatic vacua which is based on the notion of adiabatic invariants of the form  $I(t)\approx H(t)/\omega(t)$. These adiabatic invariant played an important role in the early developments of quantum mechanics and in the interpretation of the Bohr-Sommerfeld quantization condition \cite{jammer1966conceptual}. In \cite{lewis1969exact}, Lewis and Riesenfeld introduced an exact invariant $\hat{N}(t)$ which acts as a number operator and can be approximated by a series expansion up to order $\bar{n}$ in time derivatives of the frequency $\omega(t)$. The adiabatic vacuum $|0_{\bar{n}},t_0\rangle$ is then the instantaneous ground state of the adiabatic invariant of order $\bar{n}$ at the reference time $t_0$, \ie $\hat{N}(t_0)|0_{\bar{n}},t_0\rangle=0$. The Lewis-Riesenfeld invariant is related to the WKB frequency $W(t)$ by the relation
\begin{equation}
\hat{N}(t)=\frac{\hat{p}^2+\big(W(t)^2+\tfrac{\dot{W}(t)^2}{4W(t)^2}\big)\hat{q}^2+\tfrac{\dot{W}(t)}{2W(t)}(\hat{q}\hat{p}+\hat{p}\hat{q})}{2 W(t)}-\tfrac{1}{2}\,.
\label{eq:LS-intro}
\end{equation}
At the lowest order $\bar{n}=0$, the WKB frequency coincides with the instantaneous frequency $\omega(t_0)$, the adiabatic invariant $\hat{N}(t_0)$ coincides with the instantaneous number operator $\hat{a}^\dagger \hat{a}$, and the adiabatic vacuum reduces to the instantaneous vacuum. 
The adiabatic-invariants method is mostly used in the design of protocols for shortcuts to adiabaticity \cite{guery2019shortcuts} and in connection with Stokes phenomena \cite{berry1989uniform} and the optimal truncation of the adiabatic particle number
\cite{Dabrowski:2014ica,Dabrowski:2016tsx}.

\medskip

Both methods, the WKB and the adiabatic invariant construction, are adapted to the problem of a single oscillator with a time-dependent frequency, which limits their applicability to cases where one can decouple different degrees of freedom by a time-independent diagonalization. The new construction that we introduce in this paper applies to the broad class of quadratic Hamiltonians with time-dependent couplings:
\begin{align}
\hat{H}(t)=\sum_{i,j=1}^{d}& \Big(A_{ij}(t)\,\hat{p}_i \hat{p}_j+B_{ij}(t)\,\hat{q}_i \hat{q}_j\label{eq:H-gen-intro}\\[-.5em]
&\;\;\;\;\;\;+D_{ij}(t)(\hat{q}_i \hat{p}_j+\hat{p}_i \hat{q}_j)\Big)\nonumber\\[-.5em]
+\sum_{i=1}^{d}&\big(F_i(t)\, \hat{q}_i+G_i(t)\, \hat{p}_i\big)+c(t)\,.\nonumber
\end{align}
The system has $d$ bosonic degrees of freedom, an anisotropic frequency, mass matrix and magnetic coupling (coded in the time-dependent coefficients $A_{ij}(t)$, $B_{ij}(t)$ and $D_{ij}(t)$), an external driving force (coded in the time-dependent coefficients $F_{i}(t)$ and $G_{i}(t)$), and a vacuum energy depending on the coefficient $c(t)$.

Symplectic geometry and K\"ahler structures \cite{deGosson2006symplectic} provide the necessary mathematical tools for a geometric description of the dynamics defined by this class of Hamiltonians \cite{GaussianStatesFromKaehler}. The unitary time evolution operator $\hat{U}(t)$ generates linear symplectic transformations of the observables $\hat{\xi}\equiv(\hat{q}_i,\hat{p}_i)$, \ie transformations that preserve the canonical commutation relations $[\hat{q}_i,\hat{p}_j]=\ii \delta_{ij}$ and therefore the linear symplectic structure $\Omega$. In this formalism, the Hamiltonian \eqref{eq:H-gen-intro} takes the form
\begin{equation}
\hat{H}(t)=\tfrac{1}{2}\hat{\xi}\!\cdot\! h(t)\!\cdot\!  \hat{\xi}\,+\,f(t)\!\cdot  \hat{\xi}\,+c(t)\,,
\end{equation}
with the dot $\cdot$ representing matrix multiplication. The condition for the existence of a normalizable instantaneous vacuum corresponds to the requirement that the $2d\times 2d$ matrix $h(t)$ is positive definite. Using the geometric methods discussed in \cite{GaussianStatesFromKaehler}, we introduce a total number operator $\hat{N}_{J,z}$ which is defined as 
\begin{equation}
\hat{N}_{J,z}=\tfrac{1}{2}(\hat{\xi}-z)\cdot \Omega^{-1}\cdot(J-\ii \id)\cdot(\hat{\xi}-z)\,,
\label{eq:N0-def-intro}
\end{equation}
where $J$ is a linear complex structure and $z$ is a phase space vector. We then determine the matrix $J$ and the vector $z$ in terms of time derivatives of the matrix $h(t)$ and the vector $f(t)$ by requiring adiabaticity of order $\bar{n}$ in a neighborhood of a reference time $t_0$. In this way we obtain an adiabatic number operator that generalizes the Lewis-Riesenfeld invariant \eqref{eq:LS-intro} to many degrees of freedom. The adiabatic vacuum $|J,z\rangle$ is then labeled by the adiabatic complex structure $J$ and the adiabatic vector $z$, and is defined as the state with zero adiabatic particle number at the time $t_0$, \ie $\hat{N}_{J,z}|J,z\rangle=0$.

Compared to the WKB and to the Lewis-Riesenfeld construction, where one has to solve a non-linear second-order differential equation of the Ermakov kind, there is a further advantage in using the geometric construction in terms of linear complex structures: The differential equations that $J(t)$ and $z(t)$ solve are linear and first order in time. The fact that they are first order in time is a result of working with phase-space variables; linearity follows from the Hamiltonian being quadratic. The non-linearity in other approaches arises only because one uses a non-linear parametrization of the space of solutions. The linearity of the equations allows us to introduce a transparent condition of adiabaticity: We select initial conditions at a reference time $t_0$ that result in the unique solution which is analytic in an auxiliary parameter $0<\lambda<1$ that rescales time,
\begin{equation}
t\to\frac{t-t_0}{\lambda}+t_0\,,
\end{equation}
and slows down the dynamics in a neighborhood of the time $t_0$. The physical conditions of adiabaticity for the adiabatic vacuum of order $\bar{n}$ are then determined a posteriori from the condition that the expectation value of the instantaneous particle number $\hat{\mathcal{N}}_0$ in the adiabatic vacuum is small, \ie $\langle J,z|\hat{\mathcal{N}}_0|J,z\rangle\ll 1$.

\medskip

The manuscript is structured as follows: In section~\ref{sec:review}, we introduce the construction of adiabatic vacua and adiabatic particle number in terms of linear complex structures. In section~\ref{sec:adiabatic-vacua}, we describe the WKB method for a harmonic oscillator with time-dependent frequency and discuss the relation to the complex structure method. In section~\ref{sec:remarks-applicability}, we discuss the broader applicability of the complex structure method, and in section~\ref{sec:case-studies} we provide several examples. Finally in section~\ref{sec:summary}, we summarize our findings and discuss potential future work.

\section{Construction of adiabatic vacua from complex structures}\label{sec:review}

In this section we introduce the new construction of adiabatic vacua based on linear complex structures. The presentation is self-contained and is based on the geometric language of K\"ahler structures and symplectic geometry described in \cite{GaussianStatesFromKaehler}

\subsection{Quadratic Time-Dependent Hamiltonian} \label{quad_ham}
We consider a bosonic system with $d$ degrees of freedom whose classical phase space $V$ is equipped with $2d$ linear observables commonly denoted by $\xi^a\equiv(q_1,p_1,\dots,q_d,p_d)$. After quantizing the system, these observables become quantum operators $\hat{\xi}^a\equiv(\hat{q}_1,\hat{p}_1,\dots,\hat{q}_d,\hat{p}_d)$ satisfying the canonical commutation relations
\begin{align}
    [\hat{\xi}^a,\hat{\xi}^b]=\ii \Omega^{ab}\quad\text{with}\quad\Omega=\bigoplus^d_{i=1}\begin{pmatrix}
        0 & 1\\
        -1 & 0
    \end{pmatrix}\,,
    \label{eq:Omega-def}
\end{align}
where $\Omega^{ab}$ is known as the symplectic form, its inverse is denoted $\omega_{ab}$ with 
\begin{equation}
\Omega^{ac}\,\omega_{cb}=\delta^a{}_b\,,
\end{equation}
and we work in units $\hbar=1$. 
The dynamics of the system is described by an Hamiltonian $\hat{H}(t)$ which we assume to be time-dependent and quadratic in the observables $\hat{\xi}^a$:
\begin{align}
    \hat{H}(t)=\frac{1}{2}h_{ab}(t)\hat{\xi}^a\hat{\xi}^b+f_a(t)\hat{\xi}^a+c(t)\,,\label{eq:general-quadratic-H}
\end{align}
with $h_{ab}$, $f_a$ and $c$ real time-dependent coefficients. Without loss of generality we assume that $h_{ab}(t)$ is symmetric, $h_{ab}=h_{ba}$, as the antisymmetric part can be reabsorbed into the term $c(t)$. We will further assume that $h_{ab}(t)$ is a positive-definite bilinear form, which ensures that $\hat{H}(t)$ is bounded from below and has a discrete spectrum at each instant of time $t$.

Given a reference time $t_0$, the unitary evolution operator $\hat{U}(t)$ from $t_0$ to the time $t$ is given by the time-ordered exponential
\begin{equation}
\hat{U}(t)=\mathcal{T}\exp\Big[-\ii\int_{t_0}^t\hat{H}(t')dt'\Big]\,,
\label{eq:U-Texp}
\end{equation}
which solves the Schr\"odinger equation
\begin{equation}
\ii\frac{d}{dt}\hat{U}(t)=\hat{H}(t)\hat{U}(t)\,,
\label{eq:Udot}
\end{equation}
with initial condition $\hat{U}(t_0)=\id$. As the Hamiltonian takes the quadratic form \eqref{eq:general-quadratic-H}, the Heisenberg evolution of the linear observables $\hat{\xi}^a$ results in a linear symplectic transformation
\begin{equation}
\hat{U}^{-1}(t)\,\hat{\xi}^a\,\hat{U}(t)\,=\,M^a{}_b(t)\,\hat{\xi}^b+w^a(t)\,,
\label{eq:UxiU}
\end{equation}
where $M^a{}_b$ is a symplectic matrix in $\mathrm{Sp}(2d,\mathbb{R})$, \ie it preserves the symplectic form $\Omega^{ab}$,
\begin{equation}
M^a{}_c\,M^b{}_d\,\Omega^{cd}=\Omega^{ab}\,.
\end{equation}
Deriving the equation \eqref{eq:UxiU} with respect to $t$, we find that the symplectic matrix $M^a{}_b(t)$ and the translation $w^a(t)$ satisfy the equations
\begin{align}
\dot{M}(t)=&\;K(t)M(t)\,,\label{eq:Kdot}\\[.5em]
\dot{w}(t)=&\;K(t) w(t)+F(t)\,,\label{eq:wdot}
\end{align}
with initial conditions $M(t_0)=\id$ and $w(t_0)=0$. We adopt also a matrix notation $M\equiv M^a{}_b$ and denote the time derivative as $\dot{M}\equiv \frac{d}{dt}M$. The matrix $K$ and the vector $F$ are given by
\begin{align}
K^a{}_b(t)=&\;\Omega^{ac}h_{cb}(t)\,,\\[.5em]
F^a(t)=&\;\Omega^{ac}f_c(t)\,.\label{eq:driving-force}
\end{align}
Equations~\eqref{eq:Kdot} and~\eqref{eq:wdot} can be solved as a time-ordered exponential~\cite{GaussianStatesFromKaehler},
\begin{align}
M(t)&=\mathcal{T}\exp\left(\int_{t_0}^tK(t')dt'\right)M(t_0)\,,\label{eq:M(t)-exact}\\
w(t)&=M(t)w(t_0)+M(t)\int^t_{t_0}M^{-1}(t')F(t')dt'\,.
\end{align}
The matrix $K$ is an element of the Lie algebra $\mathfrak{sp}(2d,\mathbb{R})$ that generates the linear symplectic group $\mathrm{Sp}(2d,\mathbb{R})$. In this way we have encoded the unitary dynamics \eqref{eq:Udot} into the symplectic dynamics \eqref{eq:Kdot},\eqref{eq:wdot}.

\subsection{Number operator and the complex structure}
A number operator $\hat{N}_{J,z}$ for a bosonic system is defined by a choice of a linear complex structure $J^a{}_b$ and a vector $z^a$ \cite{hackl2018aspects,GaussianStatesFromKaehler}:
\begin{equation}
\hat{N}_{J,z}=\tfrac{1}{2}\omega_{ac}(J^c{}_b-\ii \delta^c{}_b)(\hat{\xi}^a-z^a)(\hat{\xi}^b-z^b)\,.
\label{eq:N-Jz-def}
\end{equation}
The linear complex structure $J^a{}_b$ is a matrix that squares to minus the identity and, together with the symplectic structure \eqref{eq:Omega-def}, defines a metric $G^{ab}$, \ie a symmetric positive-definite bilinear form:
\begin{align}
& J^a{}_c\, J^c{}_b=\,-\delta^a{}_b\,,\\[.5em]
& G^{ab}=G^{ba}\equiv -J^a{}_c\,\Omega^{cb}\,>0\,.\label{eq:G-def}
\end{align}
The operator $\hat{N}_{J,z}$ has a discrete spectrum given by the integers, with degeneracy of the eigenvalue $n\in\mathbb{N}$ given by the number of unordered partitions of $n$ into the sum of $d$ integers. 

The metric $G^{ab}$ defined by \eqref{eq:G-def} admits a physical interpretation in terms of correlations in the vacuum state of the number operator. We define the vacuum state $|J,z \rangle$ as the state satisfying
\begin{equation}
\hat{N}_{J,z}\,|J,z \rangle=0\,.
\end{equation}
As the vacuum state of a quadratic operator, this is a Gaussian state defined completely by the expectation value of linear and of quadratic bosonic operators:
\begin{align}
\langle J,z|\hat{\xi}^a|J,z \rangle=&\;z^a\,,\\[.5em]
\langle J,z|\hat{\xi}^a\hat{\xi}^b|J,z \rangle=&\;\tfrac{1}{2}(G^{ab}+
    \ii\Omega^{ab})+z^a z^b\,.
\end{align}
Higher order correlations can be expressed in terms of the metric $G^{ab}$ and the vector $z^a$ using the Wick-Isserlis theorem. In particular, this allows us to evaluate the expectation value of the operator~\eqref{eq:general-quadratic-H},
\begin{align}
    \braket{J,z|\hat{H}(t)|J,z}=&\;\tfrac{1}{2}h_{ab}(t)(\tfrac{1}{2}G^{ab}+z^az^b)\\[.5em]
    &+f_a(t)z^a+c(t)\,.\label{eq:hamiltonian expectation value}
\end{align}
which represents the average energy at the time $t$ in the ground state of the number operator $\hat{N}_{J,z}$ defined in \eqref{eq:N-Jz-def}.

\subsection{Adiabatic initial conditions}
Given a reference time $t_0$, we identify adiabatic initial conditions for the number operator $\hat{N}_0$ and its associated vacuum state $|J_0,z_0\rangle$. 

Let us consider a number operator $\hat{N}(t)$ defined by a time-dependent complex structure $J(t)$ and vector $z(t)$ as in \eqref{eq:N-Jz-def}. We require that, in a small neighborhood of the time $t_0$, this number operator defines an invariant with respect to the unitary dynamics \eqref{eq:U-Texp}, \ie
\begin{equation}
U(t)^{-1}\hat{N}(t)U(t)=\hat{N}(t_0) \,,\quad t\in (t_0-\delta,t_0+\delta)\,.
\end{equation}
Equivalently, deriving this condition with respect to $t$, we have the equation
\begin{equation}
\frac{\partial}{\partial t}\hat{N}(t)=-\ii\, [\hat{H}(t),\hat{N}(t)]\,,
\label{eq:Lewis-Riesenfield}
\end{equation}
with the initial condition $\hat{N}(t_0)=\hat{N_0}$. Here, the derivative $\frac{\partial}{\partial t}$ acts only on the time-dependent coefficients $J(t)$ and $z(t)$. This equation defines an exact invariant which generalizes the Lewis-Riesenfield invariant \cite{lewis1969exact} to a system with $d$ bosonic degrees of freedom and general quadratic time-dependent Hamiltonian. The equation by itself does not select the initial condition $\hat{N}_0$. Our goal is to determine the  initial conditions $J_0$ and $z_0$ that make the number operator $\hat{N_0}$ adiabatic in the neighborhood $t\in (t_0-\delta,t_0+\delta)$. The first step is to assume that the parameters $h_{ab}(t)$ and $f_a(t)$ in the time-dependent Hamiltonian have a slow time dependence in a neighborhood of $t_0$, which we can capture by an overall time reparametrization
\begin{equation}
t\rightarrow \frac{t-t_0}{\lambda}+t_0\,,
\end{equation}
with $\lambda$ used as a formal place-holder parameter. We require analyticity in $\lambda$ of the number operator  in the range $0<\lambda\leq 1$. For $\lambda\ll 1$, we formally slow down the dynamics and restoring $\lambda\to1$ we define the adiabatic initial condition $\hat{N}_0$.

Specifically, we assume that the number operator can be written as a formal power series in $\lambda$, 
\begin{equation}
\hat{N}^{(\lambda)}(t)=\sum_{n=0}^\infty \lambda^n\, \hat{\mathcal{N}}_n(t)\,,
\label{eq:N-eps}
\end{equation}
and solves the reparametrized equation 
\begin{equation}
\lambda \frac{\partial}{\partial t}\hat{N}^{(\lambda)}(t)=-\ii\, [\hat{H}(t),\hat{N}^{(\lambda)}]
\label{eq:Ndot-eps}
\end{equation}
order by order in $\lambda$. This requirement encodes the condition of  analyticity in $\lambda$ of the solution of the equation \eqref{eq:Ndot-eps}. The adiabatic initial condition $\hat{N}_0$ is then formally defined by setting $t=t_0$, truncating the series to the order $\bar{n}$ and restoring $\lambda=1$, 
\begin{equation}
\hat{N}_0=\sum_{n=0}^{\bar{n}}\hat{\mathcal{N}}_n(t_0)\,.
\end{equation}
Finally, the physical conditions of adiabaticity and the truncation to the order $\bar{n}$ are to be expressed in terms of physical parameters in the Hamiltonian to be determined a posteriori.

\subsection{Adiabatic initial conditions of order \texorpdfstring{$\bar{n}$}{n} for \texorpdfstring{$J_0$}{J0}, \texorpdfstring{$z_0$}{z0}}
The equations \eqref{eq:N-eps} and \eqref{eq:Ndot-eps} for the adiabatic number operator can be expressed as a set of independent conditions for the complex structure $J(t)$ and for the vector $z(t)$. 
Starting from \eqref{eq:N-Jz-def}, we define the formal power series corresponding to \eqref{eq:N-eps} as
\begin{align} 
    J^{(\lambda)}(t)=&\,\sum_{n=0}^\infty \lambda^n\,\mathcal{J}_n(t)  \,, \label{eq:ansatz_complex_strcuture}\\
    z^{(\lambda)}(t)=\,&\sum_{n=0}^\infty \lambda^n\, \zeta_n(t) \,.\label{eq:ansatz_z}
\end{align} 
The equations resulting from \eqref{eq:Ndot-eps} are
\begin{align}
     \lambda\,\dot{J}^{(\lambda)}(t)=&\,[K(t),J^{(\lambda)}(t)] \, ,\label{eq:lambdaJ-comm}\\[.5em]
   \lambda\, \dot{z}^{(\lambda)}(t)=&\,K(t) z^{(\lambda)}(t)+F(t) \,, \label{eq:complex_structure_repara}
\end{align}
to be solved order by order in $\lambda$ together with the condition $J^{(\lambda)}J^{(\lambda)}=-\id$. Hence the problem reduces to the following algebraic equations for $\mathcal{J}_n$ and $\zeta_n$:\\
for $n=0$,
\begin{align}
    [K,\mathcal{J}_0]=&\;0\,,\hspace{7.8em}\label{eq:J_0_commutator}\\[.5em]
    \mathcal{J}_0 \mathcal{J}_0=&\;-\id\,,\\[.5em]
    K\zeta_0=&\;-F\,,
\end{align}
and for $n\geq1$,
\begin{align}
    [K,\mathcal{J}_n]=&\;\dot{\mathcal{J}}_{n-1}\,, \label{eq:J_n_commutator}\\[.5em]
    \{\mathcal{J}_0,\mathcal{J}_n\}=&\;-(\mathcal{J}_1 \mathcal{J}_{n-1}+\mathcal{J}_2 \mathcal{J}_{n-2}\label{eq:J_n_anticommutator}\\
    &\qquad+\hdots+\mathcal{J}_{n-1}\mathcal{J}_1)\,, \nonumber\\
    K\zeta_n=&\;\dot{\zeta}_{n-1}\label{eq:zeta_eq}
\end{align}
where $[A,B]=AB-BA$ is the matrix commutator and $\{A,B\}=AB+BA$ the matrix anti-commutator. 

\begin{proposition}\label{prop:existence}
The linear equations ~\eqref{eq:J_0_commutator}--\,\eqref{eq:zeta_eq} satisfied by $\mathcal{J}_m$ and $\zeta_m$ have a unique solution for each $n$, provided that the coefficients $K(t)$ and $F(t)$ in the Hamiltonian are sufficiently differentiable.
\end{proposition}
\begin{proof}
We prove the proposition by induction. The base case, $n=0$, is solved by 
\begin{align}
\mathcal{J}_0=&\;|K^{-1}|\,K\,,\label{eq:J0sol}\\
\zeta_0=&\;-K^{-1}F\,.\label{eq:z0sol}
\end{align}
Note that the inverse $K^{-1}$ exists because we assumed that the Hamiltonian $\hat{H}(t)$ is bounded from below and has a discrete spectrum, and therefore $h_{ab}(t)$ is positive definite. It is also useful to note that, given a basis where $\Omega$ has the standard form \eqref{eq:Omega-def}, there exists a symplectic linear transformation preserving $\Omega$ such that the matrix $K$ is in its standard form
\begin{align}
    K\equiv\bigoplus^d_{i=1} \begin{pmatrix}
        0 & \kappa_i\\
        -\kappa_i&0
    \end{pmatrix}\,, 
    \label{eq:standard form K}
\end{align}
where $\kappa_i > 0$. As a result of Williamson's theorem~\cite{Williamson_standard_K}, for $h_{ab}>0$ it can be proven that the eigenvalues of $K$ come in complex conjugate pairs $\pm \ii \kappa_i$, which shows that $\mathcal{J}_0$ is a linear complex structure. In particular, the absolute value of the matrix $K$ can be computed using the matrix square root as $|K|=(-K K)^{1/2}$.

For $n\geq1$, one can obtain $\zeta_n(t)=K^{-1}\dot{\zeta}_{n-1}(t)$ iteratively starting from $\zeta_0(t)$, and is therefore uniquely determined in terms of time-derivatives of $K(t)$ and $F(t)$ which are assumed to be differentiable at least $n$ times. Similarly, it is possible to express $\mathcal{J}_n$ in terms of lower-order terms. First, we use the anti-commutator~\eqref{eq:J_n_anticommutator} together with~\eqref{eq:J0sol} to derive an expression for $K\mathcal{J}_n$,
\begin{align}
    K\mathcal{J}_n =&\; -|K|\big(\mathcal{J}_n |K^{-1}| K + \mathcal{J}_1 \mathcal{J}_{n-1} \\[.5em]
   &\qquad \quad+\mathcal{J}_2 \mathcal{J}_{n-2}+ \ldots + \mathcal{J}_{n-1}\mathcal{J}_1\big)\,.\nonumber 
\end{align}
We then substitute this expression into the commutator equation~\eqref{eq:J_n_commutator} and rearrange its terms to obtain an anti-commutator equation,
\begin{align}
    \{\mathcal{J}_n, |K^{-1}|\} =&\; -\big(|K^{-1}|\dot{\mathcal{J}}_{n-1} + \mathcal{J}_1 \mathcal{J}_{n-1}  \\ 
    &\quad\;\;+\mathcal{J}_2 \mathcal{J}_{n-2}+ \ldots + \mathcal{J}_{n-1}\mathcal{J}_1\big) K^{-1}\,.\nonumber 
\end{align}
This equation can be solved for $\mathcal{J}_n$ using the vectorization operation that converts an $n\times m$ matrix $A$ with entries $A_{ij}$ into the vector
\begin{equation}
\mathrm{vec}(A)=(A_{11},..,A_{m1},A_{12},..,A_{m2},..,A_{1n},..,A_{mn})^\intercal.
\end{equation}
Using this linear operation we can solve the anti-commutator equations and obtain the components of $\mathcal{J}_n$,
\begin{align}
\begin{split}
    \mathrm{vec}&(\mathcal{J}_n) = -\left((\id \otimes |K^{-1}|) + (|K^{-1}|^\intercal \otimes \id)\right)^{-1} \\
    &\qquad\qquad\quad \mathrm{vec}\big((|K^{-1}|\dot{\mathcal{J}}_{n-1} + \mathcal{J}_1 \mathcal{J}_{n-1}\\
    &\qquad\qquad\qquad\qquad+ \ldots + \mathcal{J}_{n-1}\mathcal{J}_1)K^{-1}\big)\,. \label{eq:vectorization equation}
\end{split}
\end{align}
It can be verified that~\eqref{eq:vectorization equation} is basis-independent. In an eigenbasis of $K$, the contribution from the anti-commutator, $\left((\id \otimes |K^{-1}|) + (|K^{-1}|^\intercal \otimes \id)\right)$, is diagonal and positive definite; hence, it is invertible. Furthermore, the vectorization operation is an isomorphism, so we can compute the inverse to obtain the matrix form of $\mathcal{J}_n$.
\end{proof}

The final step of the derivation of the adiabatic initial conditions is to use the formal power series \eqref{eq:ansatz_complex_strcuture} and \eqref{eq:ansatz_z} to determine $J_0$ and $z_0$. 

\begin{definition}\label{def:adiabatic-Jz}
The adiabatic initial conditions at the time $t_0$, truncated to the order $\bar{n}$, are given in terms of the evaluation at the time $t_0$ of the formal series for $\mathcal{J}_n(t)$ and $\zeta_n(t)$ defined above,
\begin{align} 
    J_0=&\,\sum_{n=0}^{\bar{n}} \,\mathcal{J}_n(t_0)\;+\mathcal{R}^{(\bar{n})}  \,, \label{eq:J0trunc}\\
    z_0=\,&\sum_{n=0}^{\bar{n}} \, \zeta_n(t_0) \,,\label{eq:z0trunc}
\end{align} 
with the remainder $\mathcal{R}^{(\bar{n})}$ enforcing the condition $J_0^{\,2}=-\id$ that $J_0$ is exactly a linear complex structure.
\end{definition}
The physical conditions under which the formal solution $J_0$ and $z_0$ represent adiabatic initial conditions for a given Hamiltonian are discussed in the next section.

\subsection{Adiabatic number \texorpdfstring{$\hat{N}_0$}{N0} and adiabatic vacuum}
Having determined the adiabatic complex structure \eqref{eq:J0trunc} and the adiabatic vector \eqref{eq:z0trunc}, we can now define a notion of adiabatic number operator $\hat{N}_0$ and adiabatic vacuum.

\begin{definition}\label{def:adiabatic-N}
Given the time-dependent Hamiltonian \eqref{eq:general-quadratic-H} and a reference time $t_0$, the adiabatic number operator $\hat{N}_0$ of order $\bar{n}$ is defined by
\begin{equation}
\hat{N}_0=\tfrac{1}{2}\omega_{ac}(J_0^{\:c}{}_b-\ii \delta^c{}_b)(\hat{\xi}^a-z_0^a)(\hat{\xi}^b-z_0^b)\,,
\label{eq:N0-def}
\end{equation}
where $J_0$ and $z_0$ are defined above in \eqref{eq:J0trunc},  \eqref{eq:z0trunc}.
\end{definition}

\begin{definition}\label{def:adiabatic-vacuum}
Given the time-dependent Hamiltonian \eqref{eq:general-quadratic-H} and a reference time $t_0$, the vacuum state of the number operator \eqref{eq:N0-def}, \ie
\begin{equation}
\hat{N}_0|J_0,z_0 \rangle=0\,,
\end{equation}
defines the adiabatic vacuum $|J_0,z_0\rangle$ of order $\bar{n}$ at the time $t_0$.
\end{definition}

The adiabatic vacuum $|J_0,z_0\rangle$ is the Gaussian state with correlation functions
\begin{align}
\langle J_0,z_0|\,\hat{\xi}^a|J_0,z_0 \rangle=&\;z_0^a\,,\\[.5em]
\langle J_0,z_0|\,\hat{\xi}^a\hat{\xi}^b|J_0,z_0 \rangle=&\;-\tfrac{1}{2}(J_0^{\,a}{}_c-
    \ii\delta^a{}_c)\Omega^{cb}+z_0^a z_0^b\,.
\end{align}
where $J_0$ and $z_0$ are defined in \eqref{eq:J0trunc},  \eqref{eq:z0trunc}. We note that, at the order $\bar{n}$, the complex structure $J_0$ and the vector $z_0$ depend on the first $\bar{n}$ time-derivatives of the functions $h_{ab}(t)$ and $f_a(t)$ appearing in the Hamiltonian \eqref{eq:general-quadratic-H}. In particular at the zero order, $\bar{n}=0$, they depend only on $h_{ab}(t_0)$ and $f_a(t_0)$ evaluated at the instant $t_0$ and not on their time derivatives. This observation leads us to the following definition.

\begin{definition}\label{def:instantaneous-vacuum}
Given the time-dependent Hamiltonian \eqref{eq:general-quadratic-H} and a reference time $t_0$, the instantaneous number operator $\hat{\mathcal{N}}_0$ and the instantaneous vacuum $|\mathcal{J}_0,\zeta_0 \rangle$ are defined by the complex structure and vector $\mathcal{J}_0$ and $\zeta_0$ obtained at the order $\bar{n}=0$ in \eqref{eq:J0sol}, \eqref{eq:z0sol}.
\end{definition}
It is useful to express the symmetrized  connected correlation function in the instantaneous vacuum in terms of the metric $\mathcal{G}^{ab}$ associated to the instantaneous complex structure $\mathcal{J}_0$:
\begin{align}
\mathcal{G}^{ab}\equiv&\;\langle \mathcal{J}_0,\zeta_0|(\hat{\xi}^a\hat{\xi}^b+\hat{\xi}^b\hat{\xi}^a)| \mathcal{J}_0,\zeta_0 \rangle-2\zeta_0^a\zeta_0^b\\[.5em]
=&\;-\big(\,|(\Omega h)^{-1}|\,\,\Omega h\Omega\big)^{ab}\,.
\end{align}
Its inverse $\mathscr{g}_{ab}$, defined by $\mathcal{G}^{ac}\mathscr{g}_{cb}=\delta^a{}_b$, satisfies the relation $\mathscr{g}=-\omega\, \mathcal{G}\,\omega$ and is given by
\begin{equation}
\mathscr{g}=|(\Omega h)^{-1}|\,h\,.
\end{equation}
Similarly, the vector $\zeta_0=-h^{-1}f$ can be expressed in terms of the parameters $h_{ab}(t_0)$ and $f_a(t_0)$ of the Hamiltonian, evaluated at the  instant $t_0$.

\medskip

With these definitions, we can now turn to the physical conditions of adiabaticity of the Hamiltonian $\hat{H}(t)$ in a neighborhood of the time $t_0$. The question can be phrased as: \emph{how close is the instantaneous vacuum to the adiabatic vacuum of order $\bar{n}$?} Concretely, the parameter that is require to be small can be taken to be the expectation value of the instantaneous number operator $\hat{\mathcal{N}}_0$ in the adiabatic vacuum of order $\bar{n}$:
\begin{equation}
 \langle J_0,z_0|\,\hat{\mathcal{N}}_0\,|J_0,z_0 \rangle\, \ll \,1\,.
 \label{eq:N-lessless-1}
\end{equation}
It is useful to write this quantity as the sum of two positive terms
\begin{align}
 &\langle J_0,z_0|\,\hat{\mathcal{N}}_0\,|J_0,z_0 \rangle=\\[.5em]
 &\qquad=-\tfrac{1}{4}\Tr(J_0\mathcal{J}_0+\id)+\tfrac{1}{2}\mathscr{g}_{ab}(z_0^a-\zeta_0^a)(z_0^b-\zeta_0^b)\\[.5em]
  &\qquad=+\tfrac{1}{8}\|J_0-\mathcal{J}_0\|^2_{\mathscr{g}}+\tfrac{1}{2}\|z_0-\zeta_0\|^2_{\mathscr{g}}\,,
\end{align}
where the Hilbert-Schmidt norm of a matrix $\|L\|^2_{\mathscr{g}}=\Tr(L L^\dagger)$ is defined in terms of the metric $\mathscr{g}$ via the adjoint  $L^\dagger= \mathcal{G} L^\intercal \mathscr{g}$. This formula implies that each of the two distances is required to be small,
\begin{equation}
\|J_0-\mathcal{J}_0\|_{\mathscr{g}}\ll1\,,\quad
\|z_0-\zeta_0\|_{\mathscr{g}}\ll1\,.
\label{eq:validity}
\end{equation}
These conditions are imposed to the truncation at each order $\bar{n}$. If the series \eqref{eq:J0trunc}, \eqref{eq:z0trunc} converges in the limit $\bar{n}\to \infty$, then we call this limit the adiabatic structure of infinite order. If the series is only asymptotic but does not converge, the question of the optimal order of truncation arises \cite{Dabrowski:2014ica,Dabrowski:2016tsx}.

\bigskip

We give some illustrative examples with simple systems with a single bosonic degree of freedom, $d=1$.

\begin{example} \label{ex_ho_1}The standard harmonic oscillator (with frequency $\omega_0>0$ and unit mass) has Hamiltonian $\hat{H}$ that is time independent,
\begin{equation}
\hat{H}=\tfrac{1}{2}\big(\hat{p}^2+\omega_0^{\,2}\,\hat{q}^2\big)\,.
\end{equation}
This Hamiltonian is defined by the positive bilinear form $h_{ab}$ which corresponds to the symplectic generator $K^a{}_b$, with
\begin{align}
    h_{ab}=\begin{pmatrix}
        \omega_0^{\,2} & 0\\[.5em]
        0 & 1
    \end{pmatrix}\,,
    \quad 
    K^a{}_b=\begin{pmatrix}
        0 & 1\\[.5em]
        -\omega_0^{\,2} & 0
    \end{pmatrix} \,.
\end{align}
The vector $F^a=0$ vanishes because there is no linear term in the Hamitonian, and therefore the adiabatic vector $z_0$ also vanishes. Moreover, as the Hamiltonian is time-independent, the adiabatic complex structure $J_0$ is independent of the reference time $t_0$ and truncates at the order $\bar{n}=0$: the instantaneous structure coincides with the structure of infinite order. Using \eqref{eq:J0sol}, we find
\begin{align}
  J_0=\begin{pmatrix}
        0 &\;\; \omega_0^{-1}\\[.5em]
        -\omega_0 & 0
    \end{pmatrix}\,,
    \qquad
  z_0=\begin{pmatrix}
        0 \\[.5em]
        0
    \end{pmatrix}\,.    
    \label{eq:ho_initial_j_g}
\end{align}
The adiabatic complex structure $J_0$ defines the adiabatic number operator
\begin{equation}
\hat{N}_0=\tfrac{1}{2\omega_0}\big(\hat{p}^2+\omega_0^2\,\hat{q}^2\big)-\tfrac{1}{2}\,,
\end{equation}
which reproduces the familiar number operators found in textbooks \cite{Sakurai:2011zz} where it is usually expressed as the normal-ordered product of a creation and an annihiliation operator, $\hat{N}_0=\hat{a}^\dagger \hat{a}$. Finally, the positive definite metric $G^{ab}$ defined by the adiabatic complex structure is
\begin{equation}
 G^{ab}\equiv\begin{pmatrix}
        \omega_0^{-1} & 0\\[.5em]
        0 & \omega_0
    \end{pmatrix}\,,
\end{equation}
which allows us to write the expectation value of quadratic operators on the adiabatic vacuum $|J_0\rangle$ as
\begin{equation}
   \tfrac{1}{2}G^{ab}\equiv \langle J_0|
    \begin{pmatrix}
        \hat{q}^2 & \frac{\hat{q}\,\hat{p}+\hat{p}\,\hat{q}}{2}\\[.5em]
         \frac{\hat{q}\,\hat{p}+\hat{p}\,\hat{q}}{2} & \hat{p}^2
    \end{pmatrix}
    |J_0 \rangle\;=
    \;\begin{pmatrix}
       \frac{1}{2\omega_0} & 0\\[.5em]
        0\;\; & \frac{1}{2}\omega_0
    \end{pmatrix}\,,
\end{equation}
which reproduces the uncertainties $(\Delta q)^2$ and $(\Delta p)^2$ given by the familiar vacuum wavefunction $\psi_0(q)=(\frac{\omega_0}{\pi})^{\frac{1}{4}}\,\ee^{-\frac{\omega_0}{2}q^2}$ in the Schr\"odinger representation.
\end{example}

\begin{example} \label{ex_ho_2}
Next, we consider a driven harmonic oscillator with time-dependent external force $\omega_0^2\,\chi(t)$,
\begin{equation}
\hat{H}(t)=\tfrac{1}{2}\big(\hat{p}^2+\omega_0^2\,\hat{q}^2\big)\,+\,\omega_0^2\,\chi(t)\,\hat{q}\,.
\end{equation}
This Hamiltonian is defined by the vector $F^a(t)=(0,-\omega_0^2\,\chi(t))$. The adiabatic vector $z_0$ of order $\bar{n}=2$ at the reference time $t_0$ is then
\begin{equation}
z_0=    \;\begin{pmatrix}
       -\chi(t_0)+\frac{\ddot{\chi}(t_0)}{\omega_0^2} \\[.7em]
        -\dot{\chi}(t_0)
    \end{pmatrix}\,,
\end{equation}
which corresponds to the expectation value
\begin{equation}
 \langle J_0,z_0|\,\hat{q}\,|J_0,z_0 \rangle\,=\, -\chi(t_0)+\frac{\ddot{\chi}(t_0)}{\omega_0^2}\,.
\end{equation}
Note that, compared to the instantaneous vacuum where the expectation value is simply the instantaneous shift $\zeta_0=-\chi(t_0)$, the adiabatic vacuum has an additional contribution that depends on times in the neighborhood of the instant $t_0$ via time-derivatives of $\chi(t)$. The physical conditions of adiabaticity can be defined in terms of the time-derivatives of $\chi(t)$ at the time $t_0$ using  dimensionless flow parameters:
\begin{equation}
\varepsilon_0(t)\equiv \frac{\dot{\chi}(t)}{\omega_0 \,\chi(t)}\,,\quad
\varepsilon_{k}(t)\equiv \frac{\dot{\varepsilon}_{k-1}(t)}{\omega_0 \,\varepsilon_{k-1}(t)}\,.
\end{equation}
Specifically, the condition \eqref{eq:N-lessless-1} at the order $\bar{n}=2$ corresponds to
\begin{align}
&\langle J_0,z_0|\,\hat{\mathcal{N}}_0\,|J_0,z_0 \rangle=\tfrac{1}{2}\|z_0-\zeta_0\|^2_{\mathscr{g}}=\frac{\dot{\chi}(t_0)^2}{2\omega_0^{\,2}}+\frac{\ddot{\chi}(t_0)^2}{2\omega_0^{\,3}}\quad\nonumber\\[.5em]
&\qquad=\,\left.\frac{\omega_0 \chi^2}{2}\varepsilon_0^{\,2}\,(1+(\varepsilon_0+\varepsilon_1)^2)\right|_{t\to t_0}\,\ll\,1\,,
\end{align}
which shows that the physical condition of adiabaticity at order $\bar{n}=2$ is $|\varepsilon_{0}(t_0)|\ll 1$ and $|\varepsilon_{1}(t_0)|\ll 1$.
\end{example}

\begin{example} \label{ex_ho_3}
Finally, we consider a simple oscillator with time-dependent frequency $\omega(t)>0$,
\begin{equation}
\hat{H}(t)=\tfrac{1}{2}\big(\hat{p}^2+\omega(t)^2\,\hat{q}^2\big)\,.\label{eq:ex_ho}
\end{equation}
At the reference time $t_0$, the instantaneous complex structure $\mathcal{J}_0$, \ie the adiabatic complex structure of order zero, coincides with the complex structure \eqref{eq:ho_initial_j_g} for the time-independent oscillator with frequency given by the value at that instant, $\omega_0=\omega(t_0)$. At the order $\bar{n}=2$ we find
\begin{align}
J_0=&\;
\begin{pmatrix}
        0&\;\; \omega^{-1}\\[.5em]
        -\omega & 0
\end{pmatrix}
\;+\;
\begin{pmatrix}
        \frac{\dot{\omega}}{2\omega^2} &0\\[.5em]
        0 & -\frac{\dot{\omega}}{2\omega^2}
\end{pmatrix}
\;+\label{eq:J-R2}\\[1em]
&+
\left.\begin{pmatrix}
            0&  \frac{\ddot{\omega}}{4\omega^{4}}-\frac{3}{8}\frac{\dot{\omega}^{2}}{\omega^{5}}\\[.5em]
                    \frac{\ddot{\omega}}{4\omega^{2}}-\frac{5}{8}\frac{\dot{\omega}^{2}}{\omega^{3}} & 0
\end{pmatrix}\right|_{t\to t_0}
\;+\mathcal{R}^{(2)}\,,\nonumber
\end{align}
where the corrections of order one and two to the instantaneus complex structure $\mathcal{J}_0$ are indicated in the first and the second line. The physical conditions of adiabaticity can be defined in terms of time-derivatives of the frequency $\omega(t)$ at the time $t_0$ using the dimensionless flow parameters
\begin{equation}
\varepsilon_0(t)\equiv \frac{\dot{\omega}(t)}{\omega(t)^2}\,,\quad
\varepsilon_{k}(t)\equiv \frac{\dot{\varepsilon}_{k-1}(t)}{\omega(t) \,\, 
\varepsilon_{k-1}(t)}\,,\label{eq:epsilon0-def}
\end{equation}
with $|\varepsilon_{0}(t_0)|\ll 1$,  $|\varepsilon_{1}(t_0)|\ll 1$ and so on. We denote $\mathcal{O}(\epsilon^n)$ terms that collectively scale as the power $n$ or higher of the flow parameters $\varepsilon_k(t_0)$, which we assume to be all of the same order  $\mathcal{O}(\epsilon)$. In particular we see that the correction of order one in \eqref{eq:J-R2} is in fact $\mathcal{O}(\epsilon)$ and the correction of order $\bar{n}=2$ is $\mathcal{O}(\epsilon^2)$. The reminder $\mathcal{R}^{(2)}$ enforces the condition the $J_0$ is a complex structure at the exact level and not just order by order. It can be obtained as a solution of the quadratic matrix equation $J_0^{\,2}=-\id$. The solution is not unique, but can be shown to be of higher order, \ie $\mathcal{R}^{(\bar{n})}=\mathcal{O}(\epsilon^{\bar{n}+1})$. For instance, we can insert the ansatz 
\begin{equation}
\mathcal{R}^{(2)}=
\begin{pmatrix}
            -r_2& 0\\[.5em]
                    0 & +r_2
\end{pmatrix}\,, \label{eq:remainder}
\end{equation}
and compute $r_2$ as the solution of the quadratic equation which is of order $\mathcal{O}(\epsilon^3)$,
\begin{equation}
r_2= \left.\Big(
 1-\sqrt{
1+
\tfrac{\ddot{\omega}}{\omega^3}
-\tfrac{15\dot{\omega}^2}{16\omega^4}
-\tfrac{\ddot{\omega}^2}{4\omega^2\,\dot{\omega}^2}
}\,\Big)
\tfrac{\dot{\omega}}{2\omega^2}\right|_{t\to t_0}\,.
\end{equation}
With these definitions, the adiabatic number operator of order $\bar{n}=2$ introduced in \eqref{eq:N0-def} evaluates to
\begin{align}
\hat{N}_0=&
\;\big(1-\tfrac{3\dot{\omega}^2}{8\omega^4}+\tfrac{\ddot{\omega}}{4\omega^3}\big)
\tfrac{\hat{p}^2}{2\omega}
+\big(1+\tfrac{5\dot{\omega}^2}{8\omega^4}-\tfrac{\ddot{\omega}}{4\omega^3}\big)
\tfrac{\omega \hat{q}^2}{2}
\\[.5em]
&\!\left.+\big(\tfrac{\dot{\omega}}{2\omega^2}-r_2\big)\tfrac{\hat{q}\hat{p}+\hat{p}\hat{q}}{2}-\tfrac{1}{2}\right|_{t\to t_0}\,.
\end{align}
Moreover we find that, for this example, the physical adiabaticity condition \eqref{eq:N-lessless-1} at $\bar{n}=2$ is 
\begin{equation}
\langle J_0,0|\,\hat{\mathcal{N}}_0\,|J_0,0 \rangle\,=\left.\frac{\dot{\omega}^2}{16\,\omega^4}\right|_{t\to t_0}\,,
\end{equation}
which is of order $\mathcal{O}(\epsilon^2)$ as expected. We report also the variance $(\Delta q)^2$ of the position operator in the adiabatic vacuum of order two,
\begin{equation}
\langle J_0,0|\,\hat{q}^2\,|J_0,0 \rangle\,=\,
        \left.\frac{1}{2\omega}+\frac{1}{8} \Big(\frac{\ddot{\omega}}{\omega^{4}}-\frac{3}{2}\frac{\dot{\omega}^{2}}{\omega^{5}}\Big)\right|_{t\to t_0}\,.
\end{equation}
In the next section we will compare the results for this example with the ones obtained using WKB methods.
\end{example}

\subsection{Adiabatic vacuum and unitary time evolution}

In discussing adiabatic vacua, it is crucial to distinguish initial conditions from dynamical equations \cite{Wigner:1985aua}. The Hamiltonian $\hat{H}(t)$ defines the unitary time evolution operator $\hat{U}(t_0,t_1)$ from the reference time $t_0$ to the time $t_1$, \eqref{eq:U-Texp}. We can also define the adiabatic number operators $\hat{N}_0$ and $\hat{N}_1$ defined in terms of adiabatic initial conditions at the reference times $t_0$ and $t_1$. Generally, the dynamics does not send adiabatic initial conditions at $t_0$ to adiabatic initial conditions at $t_1$, \ie in general
\begin{equation}
\hat{U}(t_0,t_1)^{-1}\,\hat{N}_1\,\hat{U}(t_0,t_1)\;\neq\; \hat{N}_0\,.
\end{equation} 
This fact is at the origin of particle production in dynamical backgrounds and can be expressed in terms of vacuum persistence,
or in terms of the expectation value of the number operator. Adiabaticity of the dynamics between the times $t_0$ and $t_1$ can be formulated as the condition
\begin{equation}
\langle J_0,z_0|\,\hat{U}(t_0,t_1)^{-1}\hat{N}_1\hat{U}(t_0,t_1)|J_0,z_0 \rangle\;=\;\mathcal{O}(\epsilon^{\bar{n}}) \,,
\end{equation}
for the adiabatic vacuum of order $\bar{n}$. Shortcuts to adiabaticity are protocols for the time dependence of the Hamiltonian $\hat{H}(t)$ such that the dynamics sends the adiabatic vacuum at the time $t_0$ to the adiabatic vacuum at the time $t_1$ via a non-adiabatic dynamics in the interval $t\in(t_0,t_1)$ \cite{guery2019shortcuts}. For a bosonic system with quadratic Hamiltonian \eqref{eq:H-gen-intro}, these shortcuts can be found by considering first the exact time evolution of the complex structure with adiabatic initial conditions at the time $t_0$,
\begin{align}
\tilde{J}_0(t)=&\;M(t)J_0M(t)^{-1}\,,\label{eq:MJM}\\[.5em]
\tilde{z}_0(t)=&\;M(t)z_0+M(t)\int_{t_0}^t M(t')^{-1}F(t')dt'\,,
\end{align}
and then by imposing that $\tilde{J}_0(t_1)$ and $\tilde{z}_0(t_1)$ equal exactly $J_1$ and $z_1$. We note that in the intermediate interval the unitary dynamics is well defined even if the Hamiltonian is not bounded from below. In this case there is no notion of adiabatic vacuum at the intermediate times. We illustrate some of these cases in Sec.~\ref{sec:case-studies}.

\section{Relation to WKB adiabatic vacua}\label{sec:adiabatic-vacua}

The standard construction of adiabatic vacua \cite{birrell1984quantum,fulling1989aspects,Parker:2009uva} is based on WKB methods \cite{messiah2014quantum,bender2013advanced,white2010asymptotic,Winitzki_2005} and is tailored to the case discussed in \emph{Example}~\ref{ex_ho_3} of an oscillator with time-dependent frequency. The complex structure methods introduced in Sec.~\ref{sec:review} apply to a more general class of quadratic time-dependent Hamiltonians. The goal of this section is to illustrate the relation between the two approaches.

\subsection{Constructing states from classical solutions} \label{sec:constructing_states}
We consider an oscillator with unit mass and a time-dependent frequency $\omega(t)$,
\begin{align}
    H(t)=\tfrac{1}{2}\big(p^2+\omega^2(t)\,q^2\big)\,. 
    \label{eq:time-dependant-ho}
\end{align}
The position variable $q$ obeys the classical equation of motion $\ddot{q}(t)+\omega^2(t)q(t)=0$. This linear differential equation has a $2$-dimensional space of solutions determined by the initial conditions $q(t_0)$ and $\dot{q}(t_0)$ at a reference time $t_0$. Alternatively, the $2$-dimensional space of solutions can be described as the span of two linearly independent solutions $q_1(t)$ and $q_2(t)$. The linear independence is captured by a non-vanishing Wronskian $q_1 \dot{q}_2-\dot{q}_1 q_2\neq 0$. It is useful to introduce a basis of complex-conjugate solutions $v(t)$ and $v^*(t)$ satisfying the equation of motion,
\begin{align}
    \ddot{v}(t)+\omega^2(t)v(t)=0\,,\label{eq:v-eom}
\end{align}
and the canonical Wronskian condition:
\begin{equation}
    v(t)\,\dot{v}^*(t)-\dot{v}(t)\,v^*(t)=\ii\,. \label{eq:canonical_wronskian}
\end{equation}
The equation \eqref{eq:canonical_wronskian} constrains initial conditions at the time $t_0$ and is conserved, as it can easily be checked by taking a time derivative and using \eqref{eq:v-eom}. Canonical quantization of the oscillator system consists in constructing a Hilbert space of states, together with a representation of the position and momentum operator. The Hilbert space can be defined abstractly by introducing creation and annihiliation operators $\hat{a}^\dagger$ and $\hat{a}$ with bosonic commutation relations $[\hat{a},\hat{a}^\dagger]=1$. We then introduce a reference state $|0\rangle$, the Fock vacuum, defined by the annihilation condition $\hat{a}|0\rangle=0$. The Hilbert space obtained via this Fock construction is spanned by the orthonormal basis vectors $|n\rangle=\frac{1}{\sqrt{n!}}(\hat{a}^\dagger)^n|0\rangle$. Up to this point, the vacuum state $|0\rangle$ does not have yet a physical interpretation. Finally, we introduce a representation of the position operator in the Heisenberg picture, expressed as a linear combination of the creation and annihilation operators:
\begin{equation}
\hat{q}_H(t)\equiv\hat{U}^{-1}(t)\,\hat{q}\,\hat{U}(t)\,=\,v(t)\, \hat{a} + v^*(t)\,\hat{a}^\dagger\,.
\label{eq:q_H}
\end{equation}
The position and the momentum operator, defined in the Schr\"odinger picture at the reference time $t_0$, are then given by
\begin{align}
\begin{split}
    \hat{q}&=v(t_0)\, \hat{a} + v^*(t_0)\,\hat{a}^\dagger \,,\label{eq:mode_function_definition}\\[.5em]
    \hat{p}&=\dot{v}(t_0)\,\hat{a}+\dot{v}^*(t_0)\,\hat{a}^\dagger\,.
\end{split} 
\end{align}
The canonical Wronskian condition \eqref{eq:canonical_wronskian} enforces  the canonical commutation relation $[\hat{q},\hat{p}]=\ii$ of the dynamical variables $\hat{q}=\hat{q}_H(t_0)$ and $\hat{p}=\frac{d}{dt}\hat{q}_H(t_0)$.

The representation \eqref{eq:mode_function_definition} of the position and momentum operators provides us with a physical interpretation of the state $|0\rangle$ annihilated by $\hat{a}$: The Fock vacuum now represents the Gaussian state with zero expectation value of $\hat{q}$ and $\hat{p}$ and covariance matrix
\begin{align}
       G(t_0)=&\begin{pmatrix}
        2\braket{0|\hat{q}^2|0} & \braket{0| \hat{q}\hat{p}+\hat{p}\hat{q}|0}\\[.5em]
       \braket{0|\hat{q}\hat{p}+\hat{p}\hat{q}|0} & 2\braket{0| \hat{p}^2|0}
    \end{pmatrix} \label{eq:G-v}\\[1em]
        =&\left.\begin{pmatrix}
        2|v|^2 & v\dot{v}^*+\dot{v}v^*\\[.5em]
      v\dot{v}^*+\dot{v}v^*&2|\dot{v}|^2
    \end{pmatrix}\right|_{t\to t_0}\,.\nonumber   
\end{align}
Specifically, the Gaussian state $|0\rangle$ is determined by the solution $v(t)$ of the equations \eqref{eq:v-eom} and  \eqref{eq:canonical_wronskian}, or equivalently by the initial conditions $v(t_0)$ and $\dot{v}(t_0)$ that satisfy the constraint \eqref{eq:canonical_wronskian}. The next step is to choose these initial conditions so that the Gaussian state $|0\rangle$ describes the adiabatic vacuum at a reference time $t_0$.

\subsection{WKB  adiabatic initial conditions} \label{sec:WKB_method}
The WKB construction of adiabatic vacua starts with a phase-integral parametrization of the linear equations \eqref{eq:v-eom} and \eqref{eq:canonical_wronskian}. Introducing a decomposition of $v(t)$ in an amplitude and a phase, $v(t)=A(t) \ee^{\ii\theta(t)}$, one finds that the general solution of \eqref{eq:canonical_wronskian} takes the WKB (or phase integral) form 
\begin{equation}
    v(t)=\frac{1}{\sqrt{2W(t)}}\,\exp\Bigl(-\ii\int^{t}_{t_0} W(t') \,dt'\Bigl)\,,\label{eq:para-u}
\end{equation}
and therefore a mode function $v(t)$ is completely determined by the real function $W(t)>0$, the WKB frequency. Inserting the expression \eqref{eq:para-u} into the equation of motion \eqref{eq:v-eom} we find a non-linear differential equation for $W(t)$:
\begin{equation}
     W(t)^{2}\;=\;\omega(t)^{2}-\frac{1}{2} \bigg(\frac{\ddot{W}(t)}{W(t)}-\frac{3}{2}\frac{\dot{W}(t)^{2}}{W(t)^{2}}\bigg)\,. \label{eq:dif-eq-for-W-nolambda}
\end{equation}
Let us emphasize at this point that moving from $v(t)$ to $W(t)$ is just a non-linear reparametrization and the space of solutions is still equivalent to our original $2$-dimensional space. A solution of \eqref{eq:dif-eq-for-W-nolambda} is determined by two initial conditions, $W(t_0)$ and $\dot{W}(t_0)$. 

\medskip

We turn now to the conditions of adiabaticity. Let us assume that the frequency $\omega(t)$ in the Hamiltonian \eqref{eq:time-dependant-ho} changes slowly when $t$ is in a neighborhood $(t_0-\delta,t_0+\delta)$ of the reference time $t_0$. We can capture the slow time dependence of $\omega(t)$ by an overall time reparametrization 
\begin{equation}
t\rightarrow \frac{t-t_0}{\lambda}+t_0\,,
\end{equation}
with $\lambda$ used as a formal place-holder parameter. We then assume that the function $W(t)$ can be written as a formal power series in $\lambda$,
\begin{equation}
W^{(\lambda)}(t)=\sum_{n=0}^{\infty}\lambda^n\, \mathcal{W}_n(t)
\label{eq:ansatz_W} 
\end{equation}
which solves, order by order in $\lambda$, the reparametrized equation
\begin{equation}
      W^{(\lambda)}(t)^{2}\;=\;\omega(t)^{2}-\frac{\lambda^2}{2} \bigg(\frac{\ddot{W}^{(\lambda)}(t)}{W^{(\lambda)}(t)}-\frac{3}{2}\frac{\dot{W}^{(\lambda)}(t)^{2}}{W^{(\lambda)}(t)^{2}}\bigg)\,. \label{eq:dif-eq-for-W}
\end{equation}
The condition \eqref{eq:ansatz_W} is equivalent to the requirement that $W^{(\lambda)}(t)$ is the unique solution of \eqref{eq:dif-eq-for-W} that is analytic in the parameter $\lambda$ in the range $0<\lambda\leq 1$. For $\lambda\ll 1$, we formally slow down the dynamics and restoring $\lambda\to1$ we define the adiabatic initial condition for $W^{(\lambda)}(t)$. The sequence of algebraic equations can be solved order by order finding:
\begin{align}
\mathcal{W}_0(t)=&\;\omega(t)\,,\\[.5em]
\mathcal{W}_1(t)=&\;0\,,\\
\mathcal{W}_2(t)=&\;\frac{3}{8}\frac{\dot{\omega}(t)^{2}}{\omega(t)^{3}}-\frac{\ddot{\omega}(t)}{4\,\omega(t)^2}\,,\\
\ldots\qquad&\nonumber
\end{align}
In particular, we have that the even terms $\mathcal{W}_{2n+2}(t)$ can be written in terms of time derivatives of $\mathcal{W}_{2n}(t), \mathcal{W}_{2n-2}(t),\ldots, \mathcal{W}_{2}(t), \mathcal{W}_{0}(t)$, and the odd terms vanish,  $\mathcal{W}_{2n+1}(t)=0$. 

With these notions we are now ready for the final step of the construction: We use the formal power series \eqref{eq:ansatz_W} to determine adiabatic initial conditions $W(t_0)$ and $\dot{W}(t_0)$ for the WKB frequency $W(t)$. 

\begin{definition}\label{def:adiabatic-W}
At a reference time $t_0$ and fixed a truncation order $\bar{n}$, the adiabatic initial conditions for the WKB frequency $W(t)$ are given in terms of the evaluation at the time $t_0$ of the formal series for $\mathcal{W}_n(t)$ defined above,
\begin{equation}
W_0=\sum_{n=0}^{\bar{n}} \,\mathcal{W}_n(t_0) \,, \quad
    \dot{W}_0=\sum_{n=0}^{\bar{n}} \,\dot{\mathcal{W}}_n(t_0) \,. \label{eq:W0trunc}
\end{equation}
\end{definition}
The physical conditions under which the formal solution $W_0$ and $\dot{W}_0$ represent adiabatic initial conditions for a given Hamiltonian are to be checked a posteriori, as discussed in the next section.
Specifically, at the order $\bar{n}=2$, the adiabatic initial conditions are
\begin{align}
W_0=&\;\left.\Big(\omega+\frac{3\dot{\omega}^{2}}{8\omega^{3}}-\frac{\ddot{\omega}}{4\,\omega^2}\Big)\right|_{t\to t_0}\,,
\label{eq:W0-adiab}\\[.5em]
\dot{W}_0=&\;\left.\Big(\dot{\omega}
-\frac{9\dot{\omega}^3}{8\omega^4}
+\frac{5\dot{\omega}\ddot{\omega}}{4\omega^3}
-\frac{\dddot{\omega}}{4\omega^2}
\Big)
\right|_{t\to t_0}\,.
\label{eq:W0d-adiab}
\end{align}
Equivalently, in terms of mode functions, we define:
\begin{definition}\label{def:adiabatic-v}
At a reference time $t_0$ and fixed a truncation order $\bar{n}$, the adiabatic initial conditions for the mode functions $v(t)$ are given by the value at the time $t_0$ of
\begin{align}
v_0=\frac{1}{\sqrt{2W_0}}\,,\qquad \dot{v}_0=-\frac{1}{\sqrt{2W_0}}\bigg(\frac{\dot{W}_0}{2W_0}+\ii W_0\bigg)\,,
\label{eq:v0-v0dot}
\end{align}
where $W_0$ and $\dot{W}_0$ are defined in \eqref{eq:W0trunc}.
\end{definition}
Note that the initial conditions \eqref{eq:v0-v0dot} satisfy the canonical Wronskian equation \eqref{eq:canonical_wronskian} exactly, even though $W_0$ and $\dot{W}_0$ are of order $\bar{n}$.

\begin{example}
Let us consider again the simple case of an oscillator with constant frequency $\omega(t)=\omega_0$ discussed earlier in Example~\ref{ex_ho_1}. The adiabatic initial conditions \eqref{eq:W0-adiab}--\eqref{eq:W0d-adiab} evaluate to $W_0=\omega_0$ and $\dot{W}_0=0$. With these initial conditions we can determine the solution of \eqref{eq:dif-eq-for-W-nolambda} which is simply $W(t)=\omega_0$ and the mode function
\begin{equation}
v_0(t)=\frac{1}{\sqrt{2\omega_0}}\ee^{-\ii \omega_0 \,(t-t_0)}\,,
\end{equation}
which is the familiar positive-frequency solution found in textbooks \cite{Sakurai:2011zz} and used in the definition of the Heisenberg operator $\hat{q}_H(t)$, \eqref{eq:q_H}.
\end{example}

\begin{figure}[t]
\centering
\includegraphics[width=\linewidth]{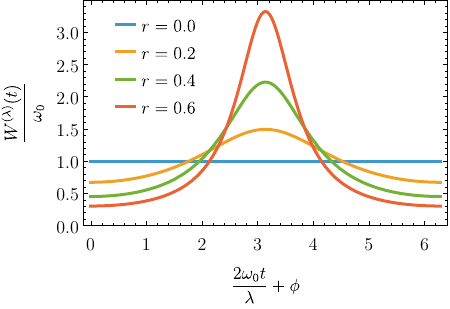}
\caption{We show the function $W^{(\lambda)}(t)$ defined in~\eqref{eq:W_harmonic_oscillator} for different choices of $r\geq0$. We see that only for $r=0$, the function is analytic in $\lambda$ and reduces to the constant $\omega_0$.}
\label{fig:W_behaviour}
\end{figure}

\textbf{Remark.} Note that the general solution of the equation \eqref{eq:dif-eq-for-W} for $\omega(t)=\omega_0$, which is the Gelfand-Dikii equation, has the form
\begin{equation}
W^{(\lambda)}(t)=\frac{\omega_0}{\cosh(2r)+\cos(\frac{2\omega_0 t}{\lambda}+\phi)\,\sinh(2r)}\,,
\label{eq:W_harmonic_oscillator}
\end{equation}
where $r>0$ and $\phi\in [0,2\pi)$ are the two parameters that identify the solution. The requirement that $W^{(\lambda)}(t)$ is an analytic function of $\lambda$, \ie it can be written as the series \eqref{eq:ansatz_W} around $\lambda=0$, imposes that $W^{(\lambda)}(t)$ is adiabatic, which happens for $r=0$. See Figure~\ref{fig:W_behaviour}.

\medskip

\textbf{Remark.} Let us highlight a subtle difference between the WKB approach discussed here and the WKB method used to solve a differential equation of the form~\eqref{eq:v-eom}, sometimes also referred to as \emph{phase integral method}, as discussed in standard textbooks on perturbation theory and asymptotic analysis~\cite{bender2013advanced,white2010asymptotic}. The phase integral method starts with an ansatz of the form
\begin{align}
    v(t)=\exp\Big(\frac{1}{\lambda}\sum_{n=0}^{\infty}\lambda^n S_n(t)\,\Big)\,.
\end{align}
This ansatz does not take into account the canonical commutation relations~\eqref{eq:canonical_wronskian} which serve as a constraint on initial data. The method outlined in section~\ref{sec:WKB_method} starts with a general solution of~\eqref{eq:canonical_wronskian} as in~\eqref{eq:para-u}, and then singular perturbation theory is applied to find solutions for the mode functions.

\subsection{Relation between the WKB method and the complex structure method}

Let us clarify first the relation between mode functions $v(t)$ and a complex structure $J_0$. The covariance matrix \eqref{eq:G-v} defines the metric $G$ compatible with the complex structure $J$, with
\begin{equation}
J_0=-G\Omega^{-1}=
\left.
\begin{pmatrix}
     -v\dot{v}^*-\dot{v}v^*    & 2|v|^2\\[.5em]
     -2|\dot{v}|^2    &v\dot{v}^*+\dot{v}v^*
    \end{pmatrix}
\right|_{t\to t_0}
\,.\label{eq:Gaussian-omega}
\end{equation}
Therefore, up to an overall phase, we can identify the state $|0\rangle$ annihilated by $\hat{a}$ with the Gaussian state $|J_0\rangle$ determined by the complex structure $J_0$,
\begin{equation}
\hat{a}\,|J_0\rangle=0\,.
\end{equation}
\textbf{Remark.} An advantage provided by the mode functions $v(t)$ is that they are solutions of the equation of motion, and therefore they provide also the time evolution of the complex structure \eqref{eq:MJM}. In fact, we have
\begin{equation}
\tilde{J}_0(t)\equiv M(t)J_0M(t)^{-1}=\mathscr{V}(t)\mathscr{W}_0\mathscr{V}(t)^{-1}
\end{equation}
with $M(t)$ the symplectic matrix satisfying the equation \eqref{eq:Kdot} and
\begin{equation}
\mathscr{V}(t)=\begin{pmatrix}
     v(t)&v^*(t)\\[.5em]
    \dot{v}(t)    & \dot{v}^*(t) 
    \end{pmatrix}\,,\quad 
\mathscr{W}_0=\begin{pmatrix}
     -\ii &0\\[.5em]
     \;\,0   & +\ii
    \end{pmatrix}.
\end{equation}
Moreover, using the phase-integral representation \eqref{eq:para-u}, we can express the initial condition for the complex structure in terms of initial conditions for the WKB frequency $W(t)$,
\begin{equation}
    J_W=
   \left.\begin{pmatrix}
        \frac{\dot{W}}{2W^2}&\frac{1}{W}\\[1em] 
       -W -\frac{\dot{W}^2}{4W^3}\;\;\;&  -\frac{\dot{W}}{2W^2}
    \end{pmatrix}
\right|_{t_0}    \,.\label{eq:equivalence_identification}
\end{equation}
We are now ready to discuss the relation between the two constructions of adiabatic vacua.
\begin{proposition}\label{prop:equivalence} For the oscillator with time-dependent frequency \eqref{eq:ex_ho}, the adiabatic complex structure $J_0$ of order $\bar{n}$ and the adiabatic WKB initial conditions $W_0$, $\dot{W}_0$ of order $\bar{n}$ define the same adiabatic vacuum, up to contributions of order $\bar{n}+1$.
\end{proposition}
\begin{proof}
Let us start with a general $2\times 2$ real matrix
 \begin{equation}
    J=
\begin{pmatrix}
       A&B\\[.5em] 
      C&  D
    \end{pmatrix}\,.
\end{equation}
The requirement that the metric $G=-J\Omega^{-1}$ is symmetric imposes $D=-A$ and the equation $J^2=-\id$ imposes then that $C=-\frac{1}{B}-\frac{A^2}{B}$. Therefore the formal series in $\lambda$ for the complex structure \eqref{eq:ansatz_complex_strcuture} takes the form
 \begin{equation}
    J^{(\lambda)}(t)=
\left.
\begin{pmatrix}
       A&B\\[.5em] 
      -\frac{1}{B}-\frac{A^2}{B}\;\;&  -A
    \end{pmatrix}
    \right|_t\,.
    \label{eq:J-lambda-A-B}
\end{equation}
The dynamic equation \eqref{eq:lambdaJ-comm} imposes then 
\begin{align}
&A(t)=-\tfrac{\lambda}{2}\, \dot{B}(t)\,,\\[.5em]
&\omega(t)^2 B(t)^2=1+\tfrac{\lambda^2}{4}\big(\dot{B}(t)^2-2B(t)\ddot{B}(t)\big)\,.
\label{eq:Bdd-lambda}
\end{align}
We note that \eqref{eq:J-lambda-A-B} has the same form as \eqref{eq:equivalence_identification} with the identification $B(t)=1/W(t)$. Moreover, with this identification, the dynamical equation \eqref{eq:Bdd-lambda} for $B(t)$ coincides with the WKB equation \eqref{eq:dif-eq-for-W} for $W^{\lambda}(t)$. Finally, the formal series expansion \eqref{eq:J0trunc} for the complex structure $J^{(\lambda)}(t)$ results in the series
\begin{equation}
B^{(\lambda)}(t)=\frac{1}{\omega(t)}+\sum_{n=1}^{\infty}\lambda^n\,\mathcal{B}_n(t)\,,
\end{equation}
which allows us to solve \eqref{eq:Bdd-lambda} order by order in $\lambda$. Finally, the truncated series at order $\bar{n}$ for $B^{(\lambda)}(t)$ and $1/W^{(\lambda)}(t)$ differ only by terms of order $\bar{n}+1$, \ie
\begin{equation}
\Big(\frac{1}{\omega}+\sum_{n=1}^{\bar{n}}\lambda^n\,\mathcal{B}_n\Big)\Big(\omega+\sum_{n=1}^{\bar{n}}\lambda^n\,\mathcal{W}_n\Big)=\mathcal{O}(\lambda^{\bar{n}+1})\,,
\end{equation}
which proves that the adiabatic complex structure $J_0$ of order $\bar{n}$ and the adiabatic WKB initial conditions $W_0$, $\dot{W}_0$ of order $\bar{n}$ define the same adiabatic vacuum, up to contributions of order $\bar{n}+1$.
\end{proof}

\medskip

To compare the two constructions, we give some examples.

\begin{example} We consider the oscillator with linearly-increasing frequency,
\begin{equation}
\omega(t)=\big(1+\tfrac{t}{\tau}\big)\,\omega_0\,.
\end{equation}
and reference time $t_0=0$. The Hamiltonian changes slowly in time  $\tau\gg \omega_0^{-1}$, as also captured by the flow parameter $\varepsilon_0$,
\begin{equation}
\varepsilon_0=\left.\frac{\dot{\omega}}{\omega^2}\right|_{t_0}=\frac{1}{\omega_0 \tau}\ll 1\,.
\end{equation}
The WKB initial conditions at order $\bar{n}=2$ are
\begin{align}
W_0=&\;\Big(1+\frac{3}{8\omega_0^2 \tau^2}\Big)\,\omega_0\\[.5em]
\dot{W}_0=&\;\Big(1-\frac{9}{8\omega_0^2 \tau^2}\Big)\,\frac{\omega_0}{\tau}
\end{align}
which result in the WKB complex structure $J_W$.
\begin{align}
J_W=\quad&
\begin{pmatrix}
       \frac{1}{2\omega_0\tau}&\frac{1}{\omega_0}-\frac{3}{8\omega_0^3 \tau^2}\\[.5em] 
    -\omega_0 -\frac{5}{8\omega_0 \tau^2}\;\;&  - \frac{1}{2\omega_0\tau}
    \end{pmatrix}+ 
    \nonumber\\[1em]
    +&
    \begin{pmatrix}
       -\frac{15}{16\omega_0^3\tau^3}&0\\[.5em] 
   0 &  \frac{15}{16\omega_0^3\tau^3}
    \end{pmatrix}\;+\;\mathcal{O}\big(\tfrac{1}{\omega_0^4\tau^4}\big)
    \,.\label{eq:JW-16}
\end{align}
This WKB complex structure can be compared to the adiabatic complex structure $J_0$ defined directly by \eqref{eq:J-R2},
\begin{align}
J_0=\quad&
\begin{pmatrix}
       \frac{1}{2\omega_0\tau}&\frac{1}{\omega_0}-\frac{3}{8\omega_0^3 \tau^2}\\[.5em] 
    -\omega_0 -\frac{5}{8\omega_0 \tau^2}\;\;&  - \frac{1}{2\omega_0\tau}
    \end{pmatrix}+ 
    \nonumber\\[1em]
    +&
    \begin{pmatrix}
       -\frac{15}{64\omega_0^3\tau^3}&0\\[.5em] 
   0 &  \frac{15}{64\omega_0^3\tau^3}
    \end{pmatrix}\;+\;\mathcal{O}\big(\tfrac{1}{\omega_0^4\tau^4}\big)
    \,.
    \label{eq:J0-64}
\end{align}
Note that $J_W$ and $J_0$ differ only at the order $\bar{n}+1=3$, \ie $\mathcal{O}\big(\tfrac{1}{\omega_0^3\tau^3}\big)$, as can be see by the coefficients $\frac{15}{16\omega_0^3\tau^3}$ and $\frac{15}{64\omega_0^3\tau^3}$ appearing in the second line of \eqref{eq:JW-16} and \eqref{eq:J0-64}.
\end{example}

\begin{example}\label{example-truncation} There are special profiles, or protocols, for the frequency $\omega(t)$ that correspond to a self-truncation of the series to a finite order $\bar{n}$. The simplest, lowest order example is the profile
\begin{equation}
\omega(t)=\frac{\omega_0}{(1-\tfrac{t}{2\tau})^2}\,.
\label{eq:omega-truncation}
\end{equation}
We consider the reference time $t_0=0$. The evolution is slow $\tau\gg \omega_0^{-1}$, \ie
\begin{equation}
\varepsilon_0=\left.\frac{\dot{\omega}}{\omega^2}\right|_{t_0}=\frac{1}{\omega_0 \tau}\ll 1\,.
\end{equation}
The WKB initial conditions of order $\bar{n}=2$ are simply
\begin{equation}
  W_0=\omega_0\,,\quad
\dot{W}_0=\frac{\omega_0}{\tau}  \,.
\end{equation}
Note that, in this case, the WKB initial conditions of order $\bar{n}=2$ and $\bar{n}>2$ coincide  (and therefore coincide trivially also with $\bar{n}\to \infty$), because of the relation $\frac{\ddot{\omega}}{\omega}-\frac{3}{2}\frac{\dot{\omega}^{2}}{\omega^{2}}=0$ that \eqref{eq:omega-truncation} satisfies. In this case we find that the  complex structure $J_W$ built via WKB methods and the adiabatic complex structure $J_0$ are identical, without any corrections of higher order, \ie $J_W=J_0$ with
\begin{equation}
J_0=
\begin{pmatrix}
       \frac{1}{2\omega_0\tau}&\frac{1}{\omega_0}\\[.5em] 
    -\omega_0 -\frac{1}{4\omega_0 \tau^2}\;\;&  - \frac{1}{2\omega_0\tau}
    \end{pmatrix}\,.
\end{equation}
This profile provides also a simple example of shortcut to adiabaticity \cite{guery2019shortcuts} as the adiabatic vacuum is of infinite order.
\end{example}


\section{Remarks on applicability}\label{sec:remarks-applicability}
In this section we comment on the applicability of the complex structure construction to phenomena where adiabatic vacua arise, such as quantum fields in cosmological spacetimes \cite{birrell1984quantum,fulling1989aspects,Parker:2009uva}.

\subsection{Time reparametrization and rescaling}
The WKB method is often used in combination with a transformation that puts the mode equation in the standard form \eqref{eq:v-eom}. The technique involves a time reparametrization together with a rescaling of the mode function, and is used for instance in the study of quantum perturbations during cosmic inflation \cite{mukhanov1992theory,Bianchi:2024qyp}  where one introduces Mukhanov-Sasaki variables~\cite{mukhanov1981quantum,sasaki1986large}. We illustrate this technique with a simple example and then show how complex structure methods apply directly to the original problem, without the need of a reparametrization. 

\medskip

We consider an oscillator with time dependent frequency $\omega(t)$ and time-dependent mass $m(t)$. The Hamiltonian of the system is
\begin{equation}
\hat{H}(t)=\frac{\hat{p}^2}{2m(t)}+\frac{1}{2}m(t)\,\omega(t)^2\,\hat{q}^2\,.
\label{eq:H-m-t}
\end{equation}
In the Heisenberg picture, the position observable $\hat{q}_H(t)$ can be expressed as 
\begin{equation}
\hat{q}_H(t)\equiv\hat{U}^{-1}(t)\,\hat{q}\,\hat{U}(t)\,=\,u(t)\, \hat{a} + u^*(t)\,\hat{a}^\dagger\,,
\end{equation}
with creation and annihilation operators defined at a reference time $t_0$ and satisfying the bosonic commutation relation $[\hat{a},\hat{a}^\dagger]=\ii$. The Heisenberg equations of motion for $\hat{q}_H(t)$ and the canonical commutation relation $[\hat{q}_H(t),\hat{p}_H(t)]=\ii$, with $\hat{p}_H(t)=m(t)\frac{d}{dt}\hat{q}_H(t)$, result in the equations
\begin{align}
& \ddot{u}(t)+\tfrac{\dot{m}(t)}{m(t)}\dot{u}(t)+\omega(t)^2\,u(t)=0\,,\label{eq:EoM-u}\\[.5em]
& u(t)\dot{u}^*(t)-\dot{u}(t)u^*(t)=\tfrac{\ii}{m(t)}\,,\label{eq:CCR-u}
\end{align}
for the mode function $u(t)$. These equations do not have the standard form \eqref{eq:v-eom}--\eqref{eq:canonical_wronskian} to which the WKB method applies. However, they can be put in the standard form via a rescaling $u(t)\to \sqrt{m(t)}\,u(t)$ of the mode function. A second transformation that one can consider is a time reparametrization $t\to \eta$ that simplifies the dependence of the frequency on the new time $\eta$. In the construction of Mukhanov-Sasaki variables, these two transformations are combined as follows.  We consider a reference time $t_0$, call $\omega_0=\omega(t_0)$, and introduce the time reparametrization $t\to \eta$ analogous to the notion of conformal time in cosmology, with
\begin{equation}
\eta(t)\equiv\;\eta_0+\int_{t_0}^t\frac{\omega(t')}{\omega(t_0)}dt'\,.
\end{equation}
We assume that this function can be inverted as $t=t(\eta)$. We denote the derivatives with respect to $t$ and to $\eta$ with $\dot{f}(t)\equiv\frac{d}{dt}f(t)$ and $g'(\eta)=\frac{d}{d\eta}g(\eta)$. We then introduce the rescaling function $z(\eta)$ defined as
\begin{equation}
\mathscr{z}(\eta)\equiv \left.\sqrt{m(t)\,\dot{\eta}(t)}\,\right|_{t\to t(\eta)}\,.
\end{equation}
The new mode function $u(t)\to v(\eta)$ is then
\begin{equation}
u(t)=\left.\tfrac{1}{\mathscr{z}(\eta)}v(\eta)\right|_{\eta\to \eta(t)}\,.
\end{equation}
In these new variables, the equations \eqref{eq:EoM-u}--\eqref{eq:CCR-u} take the standard form
\begin{align}
&v''(\eta)+\Big(\omega_0^{\,2}-\frac{\mathscr{z}''(\eta)}{\mathscr{z}(\eta)}\Big)v(\eta)=0\label{eq:v-zdd}\\[.5em]
& v(\eta)v'^*(\eta)-v'(\eta)v^*(\eta)=\ii\,. \label{eq:v-zdd-CCR}
\end{align}
We can then use the standard WKB method which applies to \eqref{eq:v-eom}--\eqref{eq:canonical_wronskian}, with the new frequency
\begin{equation}
\tilde{\omega}(\eta)\equiv\sqrt{\omega_0^2-\frac{\mathscr{z}''(\eta)}{\mathscr{z}(\eta)}}
\label{eq:omega-tilde}
\end{equation}
which defines a WKB frequency $\tilde{W}(\eta)$ and a WKB vacuum at the time $t_0$. For instance the variance of $\hat{q}$ at the time $t_0$ is 
\begin{equation}
 \langle v_0|\hat{q}^2|v_0 \rangle=\left.\frac{1}{2\,\tilde{W}(\eta)\,\mathscr{z}(\eta)^2}\right|_{\eta\to\eta(t_0)}\,.
\end{equation}
\textbf{Remark.} A time reparametrization can modify how quickly a parameter changes in the Hamiltonian. The WKB adiabatic vacuum $|v_0\rangle$ can be defined formally via the expansion of order $\bar{n}$ for $\tilde{W}(\eta)$, but the physical condition of adiabaticity is still to be determined in terms of the physical time $t$. As a related note, the frequency squared $\tilde{\omega}(\eta)^2$ defined by \eqref{eq:omega-tilde} can become negative, showing that there are non-trivial conditions for the existence of an adiabatic vacuum at the time $\eta_0$. Finally,  after finding the adiabatic vacuum in the $\eta$-time, we still have to be able to invert the transformation and obtain, at least as a series expansion, the function $\eta\to t=t(\eta)$ and the adiabatic vacuum in the $t$-time, as shown at N3LO in \cite{Bianchi:2024qyp}.

\medskip

On the other hand, the complex structure method introduced in Sec.~\ref{sec:review} applies directly to the Hamiltonian \eqref{eq:H-m-t}. One starts with the generator of symplectic transformations 
\begin{equation}
K(t)=\begin{pmatrix}
        0 & 1/m(t)\\[.5em]
        -m(t)\,\omega^2(t) & 0
    \end{pmatrix}\,.
\end{equation}
At a reference time $t_0$, the adiabatic complex structure $J_0$ is given by equation \eqref{eq:J0trunc} in terms of $K(t)$. Out of $J_0$ we can compute the correlation function $G_0=-J_0\Omega$. For instance the expectation value of the observable $\hat{q}^2$ in the adiabatic vacuum $|J_0\rangle$ at order $\bar{n}=2$ is given by the $q$-$q$-component of the matrix $\frac{1}{2}G^{ab}=-\frac{1}{2}J_0^{\,a}{}_c\Omega^{cb}$:
\begin{align}
 \langle J_0|\hat{q}^2|J_0 \rangle=
\Big(
 1&+\frac{\ddot{\omega}}{4\omega^3}-\frac{3\dot{\omega}^2}{8\omega^4}
 \label{eq:J-m-omega}\\[.5em]
 & \left.+\frac{\ddot{m}}{4m\omega^2}-\frac{\dot{m}^2}{8m^2\omega^2}
 \Big)\;\frac{1}{2m\omega}\right|_{t\to t_0}\,.\quad\nonumber
\end{align}
Note that the correlation function in the adiabatic vacuum of order $\bar{n}=2$ requires only flow parameters  $\varepsilon_n(t)$ for $\omega(t)$ and $m(t)$ up to order $\mathcal{O}(\epsilon^2)$ which, when sufficiently small, guarantie that the correlation function is positive.

\subsection{Quantum fields in cosmological spacetimes}
The complex structure methods introduced in Sec.~\ref{sec:review} apply directly to the construction of adiabatic vacua for quantum fields in cosmological spacetimes. As an example, we consider a minimally coupled massless scalar field $\varphi(\vec{x},t)$. Because of the homogeneity and isotropy of background spacetime, the action for the field decouples in Fourier modes and takes the form \cite{birrell1984quantum,fulling1989aspects,Parker:2009uva}
\begin{equation}
S[\varphi]=\!\int \!\!dt\!\int\!\!\frac{d^3 \vec{k}}{(2\pi)^2}\frac{a(t)^3}{2}\Big(|\dot{\varphi}_{\vec{k}}(t)|^2-\frac{\vec{k}^2}{a(t)^2}|\varphi_{\vec{k}}(t)|^2\Big)\,,
\label{eq:S-phi}
\end{equation}
where the scale factor $a(t)$ is defined by the Friedmann-Lema\^itre-Robertson-Walker metric, 
\begin{equation}
ds^2=-dt^2+a(t)^2\, d\vec{x}^2\,,
\end{equation}
and we have assumed flat spatial sections for simplicity. For each Fourier mode $\vec{k}$, the Hamiltonian takes exactly the form \eqref{eq:H-m-t} with time-dependent mass and frequency
\begin{equation}
m(t)=a(t)^3 \,,\qquad \omega(t)=\frac{k}{a(t)}\,,
\end{equation}
and $k\equiv |\vec{k}|$. Using \eqref{eq:J-m-omega}, we find the the adiabatic vacuum of order $\bar{n}=2$ at the time $t_0$ has equal-time correlation function
\begin{equation}
 \langle J_0|\hat{\varphi}_{\vec{k}}(t_0)\hat{\varphi}_{\vec{k}'}(t_0)|J_0 \rangle= \frac{P(k,t_0)}{k^3/2\pi^2} (2\pi)^3\delta^{(3)}(\vec{k}-\vec{k}')
\end{equation}
where $P(k,t_0)$ is the power spectrum of the adiabatic vacuum
\begin{equation}
P(k,t_0) =\Big(\frac{1}{2k\,a(t_0)^2}\,+\,\frac{2H(t_0)^2+\dot{H}(t_0)}{4k^3}\Big)\frac{k^3}{2\pi^2}
\label{eq:Pk-adiab}
\end{equation}
where $H(t)\equiv \dot{a}(t)/a(t)$ is the Hubble rate. 

On the other hand, the standard WKB technique to deal with the non-canonical kinetic term in \eqref{eq:S-phi} is to introduce Mukhanov-Sasaki variables~\cite{mukhanov1981quantum,sasaki1986large} via a reparametrization of time and a rescaling of the dynamical variable \cite{mukhanov1992theory,Bianchi:2024qyp},
\begin{equation}
t\to \eta(t)\,, \qquad \varphi_{\vec{k}}(t)\;\to\; \chi_{\vec{k}}(\eta)\equiv\,\mathscr{z}(\eta)\,\varphi_{\vec{k}}(t(\eta))
\end{equation}
with two functions $\eta(t)$ and $z(\eta)$ such that the action \eqref{eq:S-phi} takes a canonical form in the new variables. In this particular case, $\eta$ coincides with the conformal time.

\begin{example} The Bunch-Davies vacuum \cite{bunch1978quantum}.
We consider de Sitter space, the cosmological spacetime with scale factor $a_{\mathrm{dS}}(t)=\ee^{H_0 t}$ which has constant Hubble rate $H_0>0$. Using the general result \eqref{eq:Pk-adiab}, the complex structure method gives immediately the power spectrum 
\begin{equation}
P(k,t_0) =\ee^{-2H_0t_0}\frac{k^2}{4\pi^2}+\frac{H_0^2}{4\pi^2}
\label{eq:P-J-BD}
\end{equation}
for the adiabatic vacuum of order $\bar{n}=2$ at the time $t_0$. Remarkably, as it happened also in Example~\ref{example-truncation}, the expansion self-truncates at order $\bar{n}=2$ and therefore it defines also the adiabatic vacuum of infinite order $\bar{n}=\infty$. Moreover the unitary time evolution sends the adiabatic vacuum of infinite order into the adiabatic vacuum of infinite order. In fact one can check that $|J_0\rangle$ is exactly the Bunch-Davies vacuum written in terms of the complex structure \cite{bunch1978quantum}. 

On the other hand, to use the WKB technique, we determine first the relation between cosmic time $t\in (-\infty,+\infty)$ and conformal time $\eta\in (-\infty,0)$, which can be inverted in closed form. Setting $t_0=0$, $a(t_0)=1$ and $\eta_0=-\frac{1}{H_0}$, we have that the scale factor and the reparametrization scale take the form 
\begin{equation}
a_{\mathrm{dS}}(t)=\ee^{H_0 t}=-\tfrac{1}{H_0 \eta}\,,\quad\Rightarrow \quad \mathscr{z}(\eta)=-\tfrac{1}{H_0 \eta}\,.
\end{equation}
The equation \eqref{eq:v-zdd} reduces to 
\begin{equation}
v''(\eta)+\Big(k^2-\frac{2}{\eta^2}\Big)v(\eta)=0
\end{equation}
where $k=|\vec{k}|$. The WKB method allows us to define the initial conditions at the time $\eta_0$ for the frequency  $\tilde{W}(\eta)$ defined in terms of derivatives of $\tilde{\omega}(\eta)=\sqrt{k^2-\frac{2}{\eta^2}}$ at the time $\eta_0$. The series does not truncate but, in the limit $\eta_0\to-\infty$, the new frequency goes to a constant $\tilde{\omega}(\eta)\to k$, defining the WKB initial conditions for the Bunch-Davies mode function,
\begin{equation}
v_{\mathrm{BD}}(\eta)=\frac{1}{\sqrt{2k}}\Big(1-\frac{\ii}{k\eta}\Big)\,\ee^{-\ii k\eta},
\end{equation}
and the associated Bunch-Davies vacuum $|v_\mathrm{BD}\rangle$. Finally, the power spectrum at the time $t$ is obtained using the exact evolution $v(t)$ with the fixed initial conditions at $t\to-\infty$, and  is given by
\begin{align}
P(k,t)=&\;\left.\frac{|v_{\mathrm{BD}}(\eta)|^2}{\mathscr{z}(\eta)^2}\frac{k^3}{2\pi^2}\right|_{\eta\to\eta(t)}\\[.5em]
=&\;\ee^{-2H_0t}\frac{k^2}{4\pi^2}+\frac{H_0^2}{4\pi^2}\,,
\end{align}
which coincides with the result \eqref{eq:P-J-BD} obtained directlyvia complex structure methods, without the need of a reparametrization.
\end{example}



\subsection{Vacuum energy and renormalized Hamiltonian}
In the unitary time evolution of a system with time-dependent Hamiltonian, the term $c(t)$ in \eqref{eq:general-quadratic-H} contributes only an overall phase and therefore it has no effect on the state of the system. However, the physical interpretation of the Hamiltonian $\hat{H}(t)$ as the energy of the system is affected by the value of the function $c(t)$. Moreover, in the definition of the probability distribution  of work done on the system \cite{Campisi:2011wqf}, differences in energy between two different times $t_0$ and $t_1$ depend on the difference $c(t_1)-c(t_0)$. This ambiguity can be fixed by requiring the expectation value of the energy on the adiabatic vacuum to vanish. This condition fixes the function $c(t)$ in \eqref{eq:general-quadratic-H} to the value $c_{\mathrm{ren}}(t)$ defined by adiabatic subtraction, defining the renormalized Hamiltonian $\hat{H}_{\mathrm{ren}}(t)$,
\begin{align}
\hat{H}_{\mathrm{ren}}(t)=\;&\tfrac{1}{2}h_{ab}(t)\hat{\xi}^a\hat{\xi}^b+f_a(t)\hat{\xi}^a\nonumber\\[.5em]
+&\tfrac{1}{2}h_{ab}(t)(\tfrac{1}{2}J_t^{\,a}{}_c\,\Omega^{cb}-z_t^a z_t^b)-f_a(t)\,z_t^a\,.
\label{eq:H-ren}
\end{align}
In this formula, the complex structure $J_t$ and the shift $z_t$ are adiabatic of order $\bar{n}$ at the time $t$. They are defined as in \eqref{eq:J0trunc}--\eqref{eq:z0trunc} in terms of time derivatives of $h_{ab}(t)$ and $f_a(t)$ evaluated at the time $t_0=t$. Having fixed the constant $c_{\mathrm{ren}}(t)$ by adiabatic subtraction, the expectation value of the energy $\hat{H}_{\mathrm{ren}}(t)$ on the adiabatic vacuum of order $\bar{n}$ at the time $t$ vanishes by definition,
\begin{equation}
\langle J_t,z_t |\hat{H}_{\mathrm{ren}}(t)|J_t,z_t \rangle=0\,.
\end{equation}
This expression guaranties that, if we prepare the system in the adiabatic vacuum $|J_0,z_0\rangle$ of order $\bar{n}$ at the time $t_0$, the work done in the evolution from $t_0$ to $t_1$ is only of order $O(\epsilon^{\bar{n}+1})$ assuming that the flow parameters in the Hamiltonian are of order $O(\epsilon)$ in the interval $t\in (t_0,t_1)$. This requirement plays a crucial role in the definition of the renormalized energy momentum tensor in cosmological spacetimes where the adiabatic vacuum of at least order $\bar{n}=4$ is needed to avoid divergences due to the ultraviolet degrees of freedom \cite{birrell1984quantum,fulling1989aspects,Parker:2009uva,parker1974adiabatic,birrell1978application,agullo2015preferred}.

\begin{example} For the time-dependent oscillator of example~\ref{ex_ho_3}, the formula for the renormalized Hamiltonian \eqref{eq:H-ren} with $\bar{n}=2$ evaluates to
\begin{equation}
\hat{H}_{\mathrm{ren}}(t)=\tfrac{1}{2}(\hat{p}^2+\omega(t)^2\hat{q}^2)-\tfrac{\omega(t)}{2}-\tfrac{\dot{\omega}(t)^2}{16\,\omega(t)^3}\,.
\end{equation}
We note that the term $-\tfrac{\omega(t)}{2}$ corresponds to the subtraction of the instantaneous vacuum energy, which is typically enforced by the  normal ordering of creation and annihilation operators. We note that second term is of order $O(\epsilon^2)$, with the slow flow parameter for the frequency $\omega(t)$ defined in \eqref{eq:epsilon0-def}. The adiabatic vacuum of order $\bar{n}=2$ has exactly zero average energy $E_0=0$, while the instantaneous vacuum at the time $t_0$ has a small negative expectation value of the energy $\mathcal{E}_0=-\tfrac{\dot{\omega}^2}{16\,\omega^3}\big|_{t_0}$, analogous to negative energy densities in quantum field theory \cite{Ford:1996er}.
\end{example}

\section{Case studies}\label{sec:case-studies}
In this section we discuss the construction and properties of adiabatic vacua using case studies that illustrate the effect of the truncation order $\bar{n}$.

\subsection{Driven oscillator solution\texorpdfstring{: $f(t)=(A\omega_0^2\sin(\omega_1t),0)$}{}}\label{sec:driven-oscillator}
We consider the case of a driven oscillator discussed in \emph{Example}~\ref{ex_ho_2}. The Hamiltonian is of the form \eqref{eq:general-quadratic-H} with a single degree of freedom, $d=1$, constant positive-definite $h_{ab}$ and time-dependent $f_a(t)$. The driving force is $F^a(t)=\Omega^{ab}f_b(t)$,\eqref{eq:driving-force}. Specifically, we consider the case of a sinusoidally driven harmonic oscillator with $f_a(t)=(A\omega_0^2\sin(\omega_1t),0)$, which can also be interpreted as a particle in a moving trap.  The Hamiltonian of the system is 
    \begin{align}
        \hat{H}(t)&=\frac{1}{2}(\hat{p}^2+\omega_0^2 \hat{q}^2)+A\omega_0^2\sin(\omega_1 t) \,\hat{q} \,,\label{eq:H-driven}
    \end{align}
where $\omega_0$ is the frequency of the oscillator and $\omega_1$ the frequency of the driving force. The corresponding Lie algebra generator $K$ is time-independent, hence the adiabatic complex structure coincides with the complex structure $J_0$ of a time-independent harmonic oscillator with frequency $\omega_0$ as in \emph{Example}~\ref{ex_ho_1}. The adiabatic displacement vector can be determined by solving~\eqref{eq:zeta_eq} using the explicit form of $K$. In particular, we have $\zeta_0=-K^{-1}F$ and $\zeta_n=K^{-1}\dot{\zeta}_{n-1}$. Denoting $F^{(n)}$ the $n$-th derivative of $F$, we find
\begin{align}
    z_0&=-K^{-1}F(t_0)-\lambda K^{-2}\dot{F}(t_0)-\lambda^2 K^{-3}\ddot{F}(t_0)-\dots\nonumber\\
    &=-\sum^{\bar{n}}_{n=0}\lambda^nK^{-1-n}F^{(n)}(t_0)\,.\label{eq:sum-definition}
\end{align}
For the system~\eqref{eq:H-driven}, it is possible to compute the adiabatic vacuum to infinite order, $\bar{n}=\infty$, by summation of a geometric series, yielding
\begin{align}
         z_{0}=-\frac{A}{1-\lambda^2(\omega_1/\omega_0)^2}\begin{pmatrix}
             \sin(\omega_1 t_0)\\
             \lambda\omega_1\cos(\omega_1t_0)
         \end{pmatrix}\label{eq:z0-evolution}
\end{align}
where we show the expansion coefficient $\lambda$, to be set to $1$. As required, $z_0$ is an analytic function of $\lambda$. The zeroth order term in the expansion describes the instantaneous vacuum which corresponds to the particle positioned at the minimum of the quadratic potential at the time $t_0$. The next orders then encode the time dependence and show that particle is prepared with an initial velocity that compensates the movement of the trap.  In the limit $\omega_0\gg \omega_1$ the infinite order adiabatic vacuum approaches the instantaneous vacuum.

This example allows us to illustrate a non-trivial property of the adiabatic vacuum of infinite order. The exact solution $z_0$ in~\eqref{eq:z0-evolution} manifests a phenomenon that is not generic. In general, the adiabatic-following behavior is only approximate: If we prepare the adiabatic of order $\bar{n}$ at the time $t_0$, the unitary time evolution from $t_0$ to $t_1$ sends the state into an excited state that is close to the adiabatic vacuum at $t_1$, provided that the Hamiltonian changes slowly. In cases when the adiabatic vacuum of infinite order exists, as in \eqref{eq:z0-evolution}, the question arises: Does the adiabatic vacuum of infinite order evolve into the adiabatic of infinite order? In the case of the Hamiltonian \eqref{eq:H-driven} and the solution \eqref{eq:z0-evolution}, the answer is affirmative. Physically, this implies that we do not have any particle production if we prepare the system in the adiabatic vacuum of infinite order, \ie $\braket{J_0(t_1),z_0(t_1)|\hat{N}_1|J_0(t_1),z_0(t_1)}=0$ exactly, with $\hat{N}_1$ the adiabatic number operator at $t_1$ and $\ket{J_0(t_1),z_0(t_1)}=\hat{U}(t_1,t_0)\ket{J_0,z_0}$ the adiabatic vacuum at $t_0$, evolved to the time $t_1$. This behavior however is not generic, even for adiabatic vacua of infinite order, as we show explicitly later.

\subsection{Airy solution\texorpdfstring{: $\omega(t)^2=\omega_0^2+\epsilon t$}{}}

We consider now a simple oscillator with time-dependent frequency as in \emph{Example}~\ref{ex_ho_3}. Specifically we consider a frequency squared that increases linearly in time,
\begin{equation}
\omega(t)^2=\omega_0^2+\epsilon\, t\,. 
\label{eq:omega-airy}
\end{equation}
The Hamiltonian of the system is 
\begin{equation}
    \hat{H}(t)=\tfrac{1}{2}\big(\hat{p}^2+(\omega_0^2+\epsilon\, t)\,\hat{q}^2\big)\,.
\end{equation}
As in the previous case, we will determine the adiabatic vacuum of infinite order at the time $t_0$ but, in contrast to the case of the driven oscillator solution, the expansion now provides only an asymptotic expansion which does not converge.

The adiabatic complex structure $J_0$ of arbitrarily high order can be determined iteratively and expressed in terms of the WKB frequency $W(t)$ using the relation \eqref{eq:equivalence_identification}. Using the expansion \eqref{eq:W0trunc} at the reference time $t_0=0$, and expandion $1/W_0$ as a series in $\epsilon$, we find
\begin{align}
    \frac{1}{W_0}=\frac{1}{\omega_0}-\frac{5 }{32}\frac{\epsilon ^2}{\omega_0^7}+\frac{1155}{2048}\frac{\epsilon ^4}{\omega_0^{13}}-\frac{425425 }{65536}\frac{\epsilon ^6}{\omega_0^{19}}+\mathcal{O}(\epsilon^8)\,.\label{eq:W-airy-order-by-order}
\end{align}
This expansion has a convergence radius of zero, \ie the series diverges for $\epsilon>0$. Resumming the series is not immediate, but there is an elementary way of finding the function $1/W(t)$ of which this is the series in $\epsilon$ at $t_0=0$. We determine first a basis of solution of the equations of motion $\ddot{v}+\omega^2(t)v=0$. The solution space is parametrized by
\begin{align}
    v(t)=c_1\Ai(-\tfrac{t\epsilon+\omega_0^2}{\epsilon^{2/3}})+c_2\Bi(-\tfrac{t\epsilon+\omega_0^2}{\epsilon^{2/3}})\label{eq:general-sol-airy}
\end{align}
with parameters $c_1,c_2\in\mathbb{C}$ and the Airy functions $\Ai$ and $\Bi$ of the first and of the second kind. We can then use the expansion coefficients in \eqref{eq:W-airy-order-by-order} to identify the specific choice of parameters $c_1$ and $c_2$  in~\eqref{eq:general-sol-airy}, such that the expansion based on $v_0(t)$ agrees order by order with~\eqref{eq:W-airy-order-by-order}. At $t=0$, we find
\begin{align}
    v_0(t)=\frac{1}{\epsilon^{\frac{1}{6}}}\sqrt{\frac{\pi}{2}}\left[\Ai(-\tfrac{t\epsilon+\omega_0^2}{\epsilon^{2/3}})-\ii\Bi(-\tfrac{t\epsilon+\omega_0^2}{\epsilon^{2/3}})\right]\,,
\end{align}
which yields an exact expression for $W_0(t)$ via~\eqref{eq:para-u}
\begin{align}
    \frac{1}{W_0(t)}=\frac{\pi}{\epsilon^{2/3} }  \Big[\text{Ai}^2(-\tfrac{\epsilon t+\omega_0^2}{\epsilon ^{2/3}})+\text{Bi}^2(-\tfrac{\epsilon t+\omega_0^2}{\epsilon^{2/3}})\Big]\,.\label{eq:W-airy}
\end{align}
Its expansion in $\epsilon$ at $t=0$ matches~\eqref{eq:W-airy-order-by-order} order by order. Finally, we can also use $W_0(t_0)$ and $\dot{W}_0(t_0)$ to compute the adiabatic initial conditions $J_0$ via~\eqref{eq:equivalence_identification}.

\begin{figure}[t]
\centering
\includegraphics[width=\linewidth]{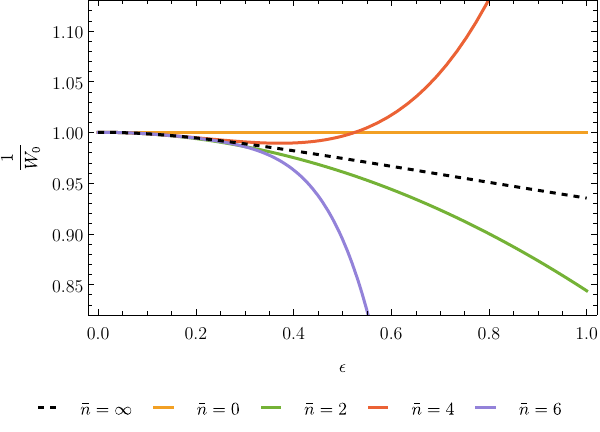}
\caption{We consider the $1/W_0$ at $t_0=0$ based on~\eqref{eq:W0trunc} for different expansion orders $\bar{n}$. The infinite order is based on the closed expression~\eqref{eq:W-airy}, while the finite orders are computed explicitly in~\eqref{eq:W-airy-order-by-order}. We see that the finite order expansions represent only asymptotic expansions, where higher orders start to divert from the infinite order solution starting from smaller values of $\epsilon$ (rather than larger ones, as for a convergent series).}
\label{fig:W vs lambda}
\end{figure}%

In figure~\ref{fig:W vs lambda}, we compare $\frac{1}{W_0}=\braket{0|\hat{q}^2|0}$ at $t_0=0$ as a function of $\epsilon$ for the first few orders $\bar{n}$ of the adiabatic expansion. It is evident that the terms in the adiabatic expansion are asymptotic to the infinite solution as $\epsilon\to 0$, but one also sees that they are only asymptotic, as higher orders start to diverge starting from smaller values of $\epsilon$ as we increase $\bar{n}$.

\begin{figure}[t]
\centering
\includegraphics[width=\linewidth]{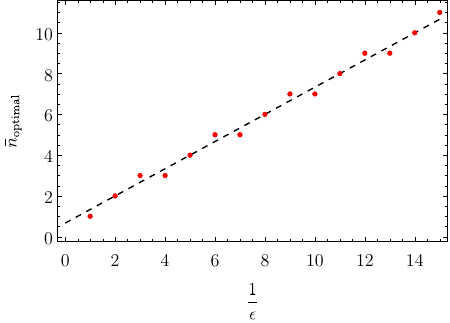}
\caption{We show for a given $1/\epsilon$ the optimal truncation order $\bar{n}_{\mathrm{optimal}}$, so that the difference between $1/W_0$ from~\eqref{eq:W-airy-order-by-order} at this order compared to the infinite order one~\eqref{eq:W-airy} is minimal. We find the asymptote $n_{\text{optimal}}\sim\frac{2}{3\epsilon}$.}
\label{fig:n-optimal}
\end{figure}

To obtain meaningful results from the expansion up to finite order, we can truncate the series at an optimal order, \ie such that for the given value of $\epsilon$ the expansion is closest to the infinite order solution. As we have the infinite order solution at $t=0$ in the present case, we can determine the optimal adiabatic order $\bar{n}_{\text{optimal}}$ by comparison. In figure~\ref{fig:n-optimal}, we show $n_{\text{optimal}}$ as a function of $1/\epsilon$. From here, we can read the behavior $n_{\text{optimal}}\sim\frac{2}{3\epsilon}$, so the optimal order increases for smaller $\epsilon$, as already deduced from figure~\ref{fig:W vs lambda}.

Clearly, the frequency $\omega(t)$ in \eqref{eq:omega-airy} will only give a positive definite Hamiltonian for $t>-\omega_0^2/\epsilon$. 
Above, we determined the  adiabatic initial condition $J_0$ to infinite order at time $t_0=0$ using the adiabatic solution \eqref{eq:W-airy}.
Can we use this solution to identify also initial condition $J_1$ to infinite order at a different time $t_1$? For this, we can reparametrize $\omega(t)$ as
\begin{align}
    \omega^2(t_1)&=\omega_0^2+\epsilon t_1
    =\omega_0^2+\epsilon (t_1-t_0)+\epsilon t_0
    =\tilde{\omega}^2_0+\epsilon t_0
\end{align}
with $\tilde{\omega}_0^2=\omega_0^2+\epsilon (t_1-t_0)$. We thus realize that we can just replace $\omega_0^2$ by $\tilde{\omega}_0^2$ in~\eqref{eq:W-airy} and evaluate at $t_0=0$ to identify the adiabatic initial conditions at time $t_1$. However, when doing so, we see immediately that this is equivalent to just evaluating $W_0(t)$ at time $t_1$ directly. Therefore, we make the same observation, as in the case of the driven oscillator solution from section~\ref{sec:driven-oscillator}, namely that the adiabatic initial condition $J_0$ of infinite order at time $t_0$ evolves into the adiabatic initial conditions $J_1$ at time $t_1$ under the unitary dynamics of the system. Again, there is no particle production if we prepare the system in the adiabatic vacuum of infinite order at the time $t_0=0$. However, for $t<-\omega_0^2/\epsilon$ there is no notion of vacuum, instantaneous or adiabatic of any order, as the system is unstable.

\subsection{Hypergeometric solution\texorpdfstring{: $\omega(t)^2=A+B\tanh(\epsilon t)$}{}}
In the previous two cases, the adiabatic initial conditions of infinite order $J_0$ at the time $t_0$ evolves \emph{exactly} into the adiabatic initial conditions $J_1$ at time $t_1$ using exact unitary time evolution. We now discuss an example that illustrates clearly that this is not the case in general.

\begin{figure}[t]
\centering
\includegraphics[width=\linewidth]{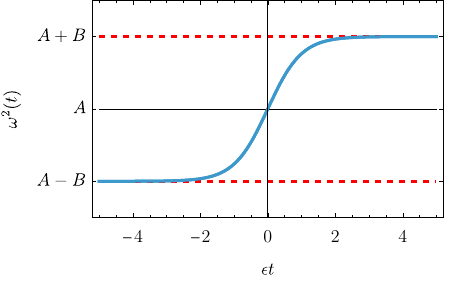}
\caption{We show the time-dependent frequency $\omega^2(t) = A + B \tanh(\epsilon t)$.}
\label{fig:tanh-potential}
\end{figure}

We consider a harmonic oscillator with frequency $\omega(t)$ that is asymptotically time-independent in the infinite past and future. The study of such systems first appeared in the context of particle creation in an expanding universe~\cite{Parker:1969au}, with an asymptotically flat spacetime in the past and in the future. Specifically, we choose a frequency profile
\begin{equation}
    \omega^2(t) = A + B \tanh(\epsilon t)\,,
\end{equation}
with $A>B$, and $\epsilon>0$  constants, as shown in figure~\ref{fig:tanh-potential} and studied in \cite{BERNARD1977201}. In the infinite past and future, the frequency takes the form
\begin{align}
    \omega_{\text{in}} &= \sqrt{A - B}\,, \quad t \to -\infty\,, \\
    \omega_{\text{out}} &= \sqrt{A + B}\,, \quad t \to +\infty\,.
\end{align}
We aim to find solutions for mode functions $v(t)$ satisfying the equation
\begin{align}
    \ddot{v}(t) + [A + B \tanh(\epsilon t)] v(t) &= 0\,. \label{eq:diff-eq-tanh}
\end{align}
Moreover, in the infinite past and future, where $\omega$ is time-independent, we require that the solutions have positive frequency and asymptotic forms:
\begin{align}
    v_{\text{in}}(t) &\sim \frac{\ee^{-\ii \omega_{\text{in}} t} }{\sqrt{2\omega_{\text{in}}}}\,, \quad v_{\text{out}}(t) \sim \frac{\ee^{-\ii \omega_{\text{out}} t} }{\sqrt{2\omega_{\text{out}}}}\,.
\end{align}
It is possible to solve~\eqref{eq:diff-eq-tanh} exactly in terms of hypergeometric functions~\cite{birrell1984quantum}, denoted by $\leftindex_2{F}_1$. The in and out solutions satisfying the positive frequency condition and correct asymptotics are given by
\begin{align}
    v_{\text{in}}(t) &= \frac{1}{\sqrt{2\omega_{\text{in}}}} \exp\{-\ii \omega_+ t - \ii \frac{\omega_-}{\epsilon} \ln[2\cosh(\epsilon t)]\} \nonumber\\
    &\hspace{-5mm}\times \leftindex_2{F}_1\left(1 + \ii \tfrac{\omega_-}{\epsilon}, \ii \tfrac{\omega_-}{\epsilon}, 1 - \ii \tfrac{\omega_{\text{in}}}{\epsilon}, \tfrac{1}{2} [1 + \tanh(\epsilon t)]\right)\,,\\[.5em]
    v_{\text{out}}(t) &= \frac{1}{\sqrt{2\omega_{\text{out}}}} \exp\{-\ii \omega_+ t - \ii \frac{\omega_-}{\epsilon} \ln[2\cosh(\epsilon t)]\} \nonumber\\
    &\hspace{-5mm}\times \leftindex_2{F}_1\left(1 + \ii \tfrac{\omega_-}{\epsilon}, \ii \tfrac{\omega_-}{\epsilon}, 1 - \ii \tfrac{\omega_{\text{out}}}{\epsilon}, \tfrac{1}{2} [1 + \tanh(\epsilon t)]\right)\,,
\end{align}
where $\omega_\pm = \frac{1}{2}(\omega_{\text{in}} \pm \omega_{\text{out}})$. 
It is evident that the in and the out solutions are not identical. However, it is possible to express the in-state as a linear combination of the out-state and its conjugate
\begin{align}
    v_{\text{in}}(t) = \alpha \,v_{\text{out}}(t) + \beta \,v_{\text{out}}^*(t)
\end{align}
with Bogoliubov coefficients $\alpha$ and $\beta$. In this case, both $\alpha$ and $\beta$ are nonzero and can be expressed in terms of Gamma functions~\cite{birrell1984quantum}.

In this setting, we only have closed expressions for the adiabatic initial conditions of infinite order at specific times, namely $J_{\mathrm{in}}$ in the infinite past and $J_{\mathrm{out}}$ in the infinite future. However, we can still use our expansion from definition~\ref{def:adiabatic-N} to define the adiabatic number operator of fixed order $\bar{n}$ and study particle production over time. We can consider
\begin{align}
    \braket{J_0(t_1),0|\hat{N}^ {(\bar{n})}_1|J_0(t_1),0}\,,\label{eq:NN}
\end{align}
where the adiabatic vacuum of infinite order at $t_0=\pm\infty$ is evolved exactly to time $t_1$, $|J_0(t_1),0\rangle$, and we evaluate the expectation value of the number operator $\hat{N}_1^{(\bar{n})}$ at $t_1$ up to order $\bar{n}$. For this, we chose the remainder $\mathcal{R}^{(\bar{n})}$ as in~\eqref{eq:remainder} yielding a quadratic equation for $r_2$, where we chose the solution that is of higher order in $\epsilon$. We can also choose the remainder $\mathcal{R}^{(2)}$ based on the WKB method of order $2$, \ie we expand $W_1$ to quadratic order and then define $\bar{N}_1$ based on~\eqref{eq:equivalence_identification}.

\begin{figure}[t]
\centering
\includegraphics[width=\linewidth]{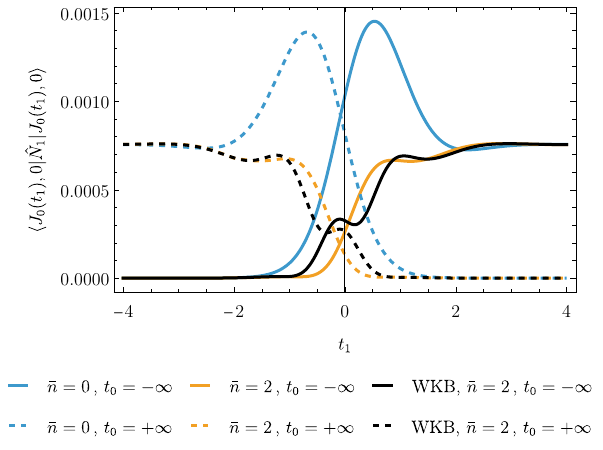}
\caption{We show $\braket{J_0(t_1),0|\hat{N}^ {(\bar{n})}_1|J_0(t_1),0}$ from~\eqref{eq:NN} for a system with $A=1$, $B=.1$ and $\epsilon=1$. For $\bar{n}>0$, this requires a choice for the remainder $\mathcal{R}^{(\bar{n})}$. For $\bar{n}=2$, we compare making this choice based on~\eqref{eq:remainder} (orange) vs. using the WKB approach (black).}
\label{fig:transition}
\end{figure}

We see in figure~\ref{fig:transition} how the particle number of one adiabatic initial condition $J_0$ at time $t_0$ with respect to the number operator $\hat{N}_{1}$ at time $t_1$ increases as we compare in and out states. This happens both for the out particle number in the in vacuum and vice versa. Note that the problem is not time symmetric, as $\omega(t)$ strictly increases going from $-\infty$ to $+\infty$. We further see that the transition shows oscillations which are due to the fact that we compute to expectation value of an adiabatic number operator of finite order  $\bar{n}=0$ and $\bar{n}=2$, instead of the adiabatic number operator of optimal order (except for the case $t_1=\pm \infty$). Finally, we note that choosing the remainder $\mathcal{R}^{(\bar{n})}$ given in~\eqref{eq:remainder}, the adiabatic vacuum based in the complex structure $J_0$ provides a smoother transition with fewer oscillations than the WKB vacuum at the same order. There are approaches to smoothen this transition further, as explained in~\cite{Dabrowski:2014ica,Dabrowski:2016tsx} based on~\cite{berry1989uniform}. In summary, this case study shows that, even if there exist adiabatic initial conditions of infinite order, such as in the limit $t_0=\pm\infty$, exact evolution in general does not evolve them into each other. This contrasts the behavior seen in the previous two case studies.

\subsection{Particle in a time-varying vertical magnetic field}
A key advantage of the construction~\ref{def:adiabatic-Jz} of adiabatic vacua in terms of complex structures is that, in contrast to the WKB method, it applies directly to systems with several degrees of freedom which can be coupled.

\begin{figure}[t]
\centering
\includegraphics[width=\linewidth]{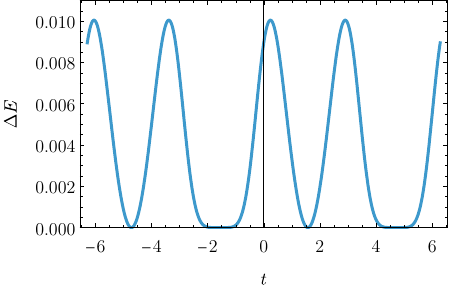}
\caption{We consider the energy difference between adiabatic vacuum $|J_0\rangle$ at order $\bar{n}=3$ and the one at order $\bar{n}=0$ (instantaneous vacuum). We choose $B(t)=B_0+B_1\sin(\lambda t)$ and $\omega=1$, $e=1$, $B_0=1$, $B_1=1$ and $\lambda=1$.}
\label{fig: energy-difference}
\end{figure}

To illustrate this feature, we consider a charged particle in an electromagnetic field, discussed also in \cite{lewis1969exact} using the Lewis-Riesenfeld invariant. The system consists of a $3$ dimensional isotropic oscillator with time-independent frequency $\omega$, coupled to a time-dependent magnetic field $\vec{B}=(0,0,B(t))$ in the $z$-direction. The Hamiltonian of the system is 
\begin{align}
    H=\tfrac{1}{2}\big(\vec{p}+e\vec{A}(t)\big)^{2}+\tfrac{1}{2}\omega^{2}\vec{q}\,{}^2\,,
\end{align}
where $e$ is the charge of the particle. We choose the gauge $\vec{A}=\frac{1}{2}(-yB,xB,0)$ for the vector potential, resulting in the gauge-fixed quadratic Hamiltonian
\begin{align}
\begin{split}
    \hat{H}(t)={}&\frac{1}{2}(\hat{p}_{x}^{2}+\hat{p}_{y}^{2}+\hat{p}_{z}^{2})+e B(t)(\hat{x}\hat{p}_{y}-\hat{y}\hat{p}_{x})\\
    &+\big(\omega^{2}+\tfrac{1}{4}e^2 B(t)^{2}\big)(\hat{x}^{2}+\hat{y}^{2})+\omega^{2}\hat{z}^{2}\,.
\end{split}
\end{align}
We compute the instantaneous and the adiabatic vacuum up to second order. As the subsystem $(z,p_z)$ is decoupled and time-independent, it suffices to compute the adiabatic vacuum for the subsystem $(\hat{x},\hat{p}_x,\hat{y},\hat{p}_y)$. We then write $h_{ab}$ as
\begin{align}
    h = 
        \begin{pmatrix}
            \omega^{2} + \frac{e^2B^{2}}{4} & 0 & 0 & \frac{eB}{2} \\
            0 & \omega^{2} + \frac{e^2B^{2}}{4} & -\frac{eB}{2} & 0 \\
            0 & -\frac{eB}{2} & 1 & 0 \\
            \frac{eB}{2} & 0 & 0 & 1 \\
        \end{pmatrix}\,.
\end{align}
We compute the expansion of the complex structure $J_0$ up to order $\bar{n}=2m+1$ as
\begin{align}
    J_0=\sum^{m}_{n=0}\lambda^{2n}\begin{pmatrix}
        \lambda X_{2n+1}  & 0 & X_{2n} & 0\\
        0 & \lambda X_{2n+1} & 0 & X_{2n}\\
        Y_{2n} & 0 & -\lambda X_{2n+1} & 0\\
        0 & Y_{2n} & 0 & -\lambda X_{2n+1}
    \end{pmatrix}\,,
\end{align}
with the coefficients $X_n$ and $Y_n$ given by: at order $0$, $1$,
\begin{align}
    X_0&=\frac{2}{\sqrt{4\omega^{2}+e^2B^{2}}}\,,&
    Y_0&=-\frac{\sqrt{4\omega^{2}+e^2B^{2}}}{2}\,,\\
    X_1&=\frac{e^2B\dot{B}}{(4\omega^{2}+e^2B^{2})^{\frac{3}{2}}}\,,
\end{align}    
at order $2$,
\begin{align}
    X_2&=\frac {(8 \omega^{2} - 3 e^2B^{2})e^2\dot {B}^{2} + 
        2 (4 \omega^{2} + e^2B^{2})e^2B \ddot {B}} {(4 \omega^{2} + 
        e^2B^{2})^{\frac {7} {2}} }\,,\\
    Y_2&=\frac {(8 \omega^{2} - 5 e^2B^{2})e^2\dot {B}^{2} + 
         2 (4 \omega^{2} + e^2B^{2})e^2 B\ddot {B}} {4 (4 \omega^{2} + 
         e^2B^{2})^{\frac{5}{2}} }\,,
\end{align}
and at order $3$,
\begin{widetext}
\begin{align}
\begin{split}
    X_3=\frac { 
   e \dot {B} (2 e\ddot {B} (e^2 B^2 + 
          4 \omega^2 ) (-12\omega^2 + 7 e^2 B^2 ) + 
      5 e^3 B \dot {B}^2 (16 \omega^2-3 e^2 B^2 ) )-2 e^2 B \dddot {B}(4 \omega^2+e^2 B^2 )^2 } {2 (4\omega^2 + e^2 B^2)^{9/2}}\,.
\end{split}
\end{align}
\end{widetext}
It is interesting to compute the energy difference $\Delta E$ between the adiabatic vacuum of order $\bar{n}=3$ and instantaneous vacuum at the same time. We observe that there is a finite difference that depends on how slowly the magnetic field changes. As shown in figure~\ref{fig: energy-difference}, the energy difference $\Delta E$ is positive, as at each time $t_0$ there is no state with a smaller energy than the instantaneous vacuum. We further notice that there are specific times where the energy difference vanishes, implying that the adiabatic vacuum of order $\bar{n}=3$ and the instantaneous vacuum coincide at these instants.

\section{Discussion}\label{sec:summary}

We summarize the results presented in this paper and discuss directions for further investigations.

In this work, we introduced a new construction of adiabatic initial conditions for time-dependent quadratic Hamiltonians in bosonic systems. The construction is based on the complex structure formalism for Gaussian states~\cite{GaussianStatesFromKaehler} and specifies a pair $(J_0,z_0)$ consisting of a complex structure $J_0$ and a displacement vector $z_0$ via a formal power series (see  \emph{Definition}~\ref{def:adiabatic-Jz}). This structure gives rise to both an adiabatic number operator $\hat{N}_0$ (see \emph{Definition}~\ref{def:adiabatic-N}) and an adiabatic vacuum (see \emph{Definition}~\ref{def:adiabatic-vacuum}) as its ground state. We showed that the defining condition results in well-defined framework with a unique solution and we provided an algorithm to compute the adiabatic vacuum order by order using a simple recursion relation, which involves only  derivatives of the matrix that defines the Hamiltonian of the system (see \emph{Proposition}~\ref{prop:existence}). Moreover, we identified the physical conditions of adiabaticity of order $\bar{n}$ in terms of the time-dependence of parameters of the Hamiltonian (See Eq.~\eqref{eq:N-lessless-1}--\eqref{eq:validity})

To illustrate the broad applicability of the new complex structure methods for adiabatic vacua, we compared them to the more familiar WKB construction which is adapted to a single degree of freedom, $d=1$. Specifically, we showed that the two methods, the complex structure method and the WKB method, yield the same adiabatic vacuum state (see \emph{Proposition}~\ref{prop:equivalence}), whenever we can apply the latter. We noted also that the equivalence is up to higher order corrections in the $\bar{n}$ truncation, and the linear complex structure methods brings in the technical advantage that all recursive equations are linear. The complex structure method thus provides a~\emph{natural generalization} that goes beyond systems with the simple Hamiltonian $\hat{H}(t)=\frac{1}{2}(\hat{p}^2+\omega(t)^2\hat{q}^2)$. Finally, we discussed the applicability of the complex structure method in the context of cosmological spacetimes and vacuum energy renormalization and illustrated it in several case studies.


\medskip

The results of this paper open up a number of new directions for future investigation. We describe some of them briefly here. 

The techniques we introduced focused on a bosonic system with a quadratic Hamiltonian. Using the K\"ahler structure methods of \cite{GaussianStatesFromKaehler}, the results on adiabatic vacua can be generalized to fermionic systems with quadratic Hamiltonian, and thus indirectly also to those spin chain Hamiltonians that can be mapped to free fermions via the Jordan-Wigner transformation \cite{coleman2015introduction}. We plan to investigate this direction and determine a notion of adiabatic vacua for bosonic and fermionic systems defined in terms of adiabatic K\"ahler structures.

In this paper we focused on adiabatic vacua for systems with a finite number $d$ of bosonic degrees of freedom, and it would be interesting to study various aspects that arise in the limit $d\to\infty$ of infinitely many degrees of freedom. Complex structures have been used for defining the Fock space of quantum fields \cite{Ashtekar:1975zn,Ashtekar:1980yw,Wald:1995yp,Derezinski:2013dra} and the issue of unitarily inequivalent representations of the field observables in Fock space \cite{Agullo:2015qqa, Much:2018ehc, Cortez:2019orm}, but not as a tool for determining adiabatic vacua \cite{Parker:1968mv,Parker:1969au,Parker:2012at}. It would be interesting to extend and use the methods introduced in this paper to investigate questions on adiabatic initial states of quantum fields \cite{Ashtekar:2016pqn,Handley:2016ods,Fahn:2018ahm,ElizagaNavascues:2020fai,Martin-Benito:2021szh,Bianchi:2024jmn} and the renormalization of the energy-momentum tensor \cite{Negro:2024bbf,Kolb:2023ydq,Animali:2022lig,Ferreiro:2023uvr} in curved spacetimes, or in the presence of time-dependent external sources, using adiabatic complex structures.

Moreover, the limit $d\to\infty$ is relevant for many-body systems and quantum fields on a lattice, where one studies a ``thermodynamic limit'' given by many lattice sites at fixed density. Complex structure methods have been used in studies of entanglement in lattice systems \cite{Bianchi:2015fra,Bianchi:2017kgb}, in loop quantum gravity \cite{Bianchi:2016hmk,Ashtekar:2021kfp} and in spin chains \cite{Vidmar:2017uux,Vidmar:2018rqk} and, using the methods introduced in this paper, the construction can be extended to adiabatic vacua and the renormalization of observables.

For finitely many defrees of freedom, vacuum energy is generally defined in terms of the instantaneous vacuum. For instance, for $d$ oscillators with frequency $\omega_k(t)$, the vacuum energy is given by $E_0(t_0)=\sum_{k=1}^d\frac{1}{2}\hbar \omega_k(t_0)$. If this was the correct physical value, then the vacuum energy would have a large backreaction for sufficiently large $d$ \cite{Bianchi:2010uw,bianchi2010dark}. Another standard prescription consists in subtracting the instantaneous vacuum energy, and defining the energy of a state as the difference with respect to the instantaneous vacuum. However, when the frequency changes in time, for large $d$ the difference $E_0(t_1)-E_0(t_0)$ can be large. In this paper we introduced a renormalized Hamiltonian defined via adiabatic subtraction \eqref{eq:H-ren}. The construction follows the techniques for the renormalization of quantum fields \cite{parker1974adiabatic,birrell1978application,agullo2015preferred} and generalizes it to complex structures in many-body systems. By construction, using the adiabatic vacuum of order $\bar{n}$, the vacuum energy difference $E^{(\bar{n})}_0(t_1)-E^{(\bar{n})}_0(t_0)$ can be small for slowly changing frequency and large $d$. It would be interesting to investigate its effect on work fluctuation relations and its measurability in experiments \cite{Campisi:2011wqf}.

Finally, in the case of a single degree of freedom $d=1$, there are significant results on the convergence of the asymptotic expansion of order $\bar{n}$ for adiabatic vacua and the adiabatic number operator, with the identification of the optimal order of truncation \cite{Dabrowski:2014ica,Dabrowski:2016tsx} and the relation to Stokes phenomena \cite{berry1989uniform}. It would be interesting to explore the generalization of these results to $d$ degrees of freedom using the complex structure methods for adiabatic vacua introduced in this paper.

\medskip

\begin{acknowledgements}
We thank Abhay Ashtekar, Miguel Fernandez Flores and Jorge Sofo for useful discussions. E.B. acknowledges support from the National Science Foundation, Grant No. PHY-2207851. This work was made possible through the support of the ID 62312 grant from the John Templeton Foundation, as part of the project \href{https://www.templeton.org/grant/the-quantum-information-structure-of-spacetime-qiss-second-phase}{``The Quantum Information Structure of Spacetime'' (QISS)}. The opinions expressed in this work are those of the authors and do not necessarily reflect the views of the John Templeton Foundation.
\end{acknowledgements}


%


\end{document}